\documentclass[aps,prxquantum,twocolumn,superscriptaddress,floatfix,longbibliography,letter]{revtex4-1}
\usepackage[bookmarks=true,colorlinks,linkcolor=OrangeRed,urlcolor=NavyBlue,citecolor=RoyalBlue]{hyperref}
\usepackage{epsfig,amsmath,amssymb,color,dsfont,upgreek,physics}
\usepackage{mathrsfs}
\usepackage{mathtools}
\usepackage{bbold}
\usepackage[makeroom]{cancel}
\usepackage[caption=false]{subfig}
\usepackage{mathrsfs}
\usepackage{graphicx}
\usepackage{subfig}
\usepackage[countmax]{subfloat}
\usepackage[english]{babel}
\usepackage[dvipsnames]{xcolor}

\usepackage[normalem]{ulem}

\usepackage{braket}

\definecolor{mypurple}{rgb}{0.49,0.18,0.56}
\definecolor{mygold}{rgb}{0.93,0.69,0.13}
\definecolor{mygreen}{rgb}{0,0.5,0}
\definecolor{myblue}{rgb}{0,0,0.75}
\definecolor{mymagenta}{cmyk}{0,1,0,0.12}
\definecolor{mygray}{rgb}{0.5,0.5,0.5}

\usepackage{umoline}

\definecolor{mypink1}{rgb}{0.858, 0.188, 0.478}

\begin{document}

\title{Enhancing disorder-free localization through dynamically emergent local symmetries}
\author{Jad C.~Halimeh}
\email{jad.halimeh@physik.lmu.de}
\affiliation{INO-CNR BEC Center and Department of Physics, University of Trento, Via Sommarive 14, I-38123 Trento, Italy}
\author{Lukas Homeier}
\affiliation{Department of Physics and Arnold Sommerfeld Center for Theoretical Physics (ASC), Ludwig-Maximilians-Universit\"at M\"unchen, Theresienstra\ss e 37, D-80333 M\"unchen, Germany}
\affiliation{Munich Center for Quantum Science and Technology (MCQST), Schellingstra\ss e 4, D-80799 M\"unchen, Germany}
\affiliation{Department of Physics, Harvard University, Cambridge, Massachusetts 02138, USA}
\author{Hongzheng Zhao}
\affiliation{Blackett Laboratory, Imperial College London, London SW7 2AZ, United Kingdom}
\affiliation{Max-Planck-Institut f\"ur Physik komplexer Systeme, N\"othnitzer Stra\ss e 38, 01187 Dresden, Germany}
\author{Annabelle Bohrdt}
\affiliation{ITAMP, Harvard-Smithsonian Center for Astrophysics, Cambridge, MA 02138, USA}
\affiliation{Department of Physics, Harvard University, Cambridge, Massachusetts 02138, USA}
\author{Fabian Grusdt}
\affiliation{Department of Physics and Arnold Sommerfeld Center for Theoretical Physics (ASC), Ludwig-Maximilians-Universit\"at M\"unchen, Theresienstra\ss e 37, D-80333 M\"unchen, Germany}
\affiliation{Munich Center for Quantum Science and Technology (MCQST), Schellingstra\ss e 4, D-80799 M\"unchen, Germany}
\author{Philipp Hauke}
\affiliation{INO-CNR BEC Center and Department of Physics, University of Trento, Via Sommarive 14, I-38123 Trento, Italy}
\author{Johannes Knolle}
\affiliation{Department of Physics, Technische Universit\"at M\"unchen, James-Franck-Straße 1, D-85748 Garching, Germany}
\affiliation{Munich Center for Quantum Science and Technology (MCQST), Schellingstra\ss e 4, D-80799 M\"unchen, Germany}
\affiliation{Blackett Laboratory, Imperial College London, London SW7 2AZ, United Kingdom}

\begin{abstract}
Disorder-free localization is a recently discovered phenomenon of nonergodicity that can emerge in quantum many-body systems hosting gauge symmetries when the initial state is prepared in a superposition of gauge superselection sectors. Thermalization is then prevented up to all accessible evolution times despite the model being nonintegrable and translation-invariant. In a recent work [Halimeh, Zhao, Hauke, and Knolle,~\href{https://arxiv.org/abs/2111.02427}{arXiv:2111.02427}], it has been shown that terms linear in the gauge-symmetry generator stabilize disorder-free localization in $\mathrm{U}(1)$ gauge theories against gauge errors that couple different superselection sectors. Here, we show in the case of $\mathbb{Z}_2$ gauge theories that disorder-free localization can not only be stabilized, but also \textit{enhanced} by the addition of translation-invariant terms linear in a local $\mathbb{Z}_2$ \textit{pseudogenerator} that acts identically to the full generator in a single superselection sector, but not necessarily outside of it. We show analytically and numerically how this leads through the quantum Zeno effect to the dynamical emergence of a renormalized gauge theory with an enhanced local symmetry, which contains the $\mathbb{Z}_2$ gauge symmetry of the ideal model, associated with the $\mathbb{Z}_2$ pseudogenerator. The resulting proliferation of superselection sectors due to this dynamically emergent gauge theory creates an effective disorder greater than that in the original model, thereby enhancing disorder-free localization. We demonstrate the experimental feasibility of the $\mathbb{Z}_2$ pseudogenerator by providing a detailed readily implementable experimental proposal for the observation of disorder-free localization in a Rydberg setup.
\end{abstract}
\date{\today}
\maketitle
\tableofcontents

\section{Introduction}
Gauge theories are a powerful framework allowing the description of the laws of nature through local constraints that must be satisfied at every point in space and time \cite{Weinberg_book}. They are fundamental in the description of interactions between elementary particles as mediated through gauge bosons, and dictate an intrinsic relation between charged matter and the surrounding electromagnetic field \cite{Gattringer_book}. In quantum electrodynamics, this is manifest in the famed Gauss's law, which protects salient features such as a massless photon and a long-ranged Coulomb law \cite{Zee_book}.

The crucial property of a gauge theory $\hat{H}_0$ is its underlying gauge symmetry, which defines an extensive set of local constraints. Restricting our discussion to Abelian gauge theories, the gauge symmetry has a generator $\hat{G}_j$ defined at the site $j$, where matter fields reside, and its adjacent links, where gauge fields lie. Gauge invariance is embodied in the commutation relation $\big[\hat{H}_0,\hat{G}_j\big]=0,\,\forall j$. The eigenvalues $g_j$ of $\hat{G}_j$ are so-called background charges, and a set of them over the entire volume of the system defines a gauge-invariant superselection sector. In recent years, lattice gauge theories \cite{Kogut1975,Rothe_book} have witnessed considerable effort in their experimental realization in low-energy table-top setups of quantum synthetic matter \cite{Martinez2016,Muschik2017,Bernien2017,Klco2018,Kokail2019,Lu2019,Goerg2019,Schweizer2019,Mil2020,Klco2020,Yang2020,Zhou2021}. These experiments have allowed for exploring fundamental high-energy physics phenomena, such as the particle--antiparticle creation \cite{Martinez2016}, Coleman's phase transition \cite{Coleman1976,Bernien2017,Kokail2019,Yang2020}, and the thermalization dynamics of gauge theories \cite{Zhou2021}, to name a few, and they hold the promise of probing exotic far-from-equilibrium behavior in strongly-correlated matter \cite{Dalmonte_review,aidelsburger2021cold,Zohar_review}.

Recent research in far-from-equilibrium quantum many-body systems has unraveled intriguing phenomena in their relaxation dynamics \cite{Mori_review}. Whereas quench dynamics propagated by integrable models will not thermalize but instead relax to a generalized Gibbs ensemble arising from the plethora of local integrals of motion \cite{Rigol2006,Rigol2007}, generic nonintegrable systems are expected to thermalize according to the eigenstate thermalization hypothesis (ETH) \cite{Rigol_review}. In the presence of quenched disorder in interacting models, many-body localization (disorder-MBL) arises \cite{Basko2006}, which violates ETH and leads to localized dynamics in local observables. Disorder-MBL has been the subject of intense theoretical investigation recently \cite{Alet_review,Abanin_review} and has been experimentally probed in various quantum synthetic matter setups \cite{Schreiber2015,Kondov2015,Smith2016,Choi2016,Roushan2017,chiaro2020direct,Rispoli2019,Lukin2019}.

\begin{figure}[t!]
	\centering
	\includegraphics[width=.5\textwidth]{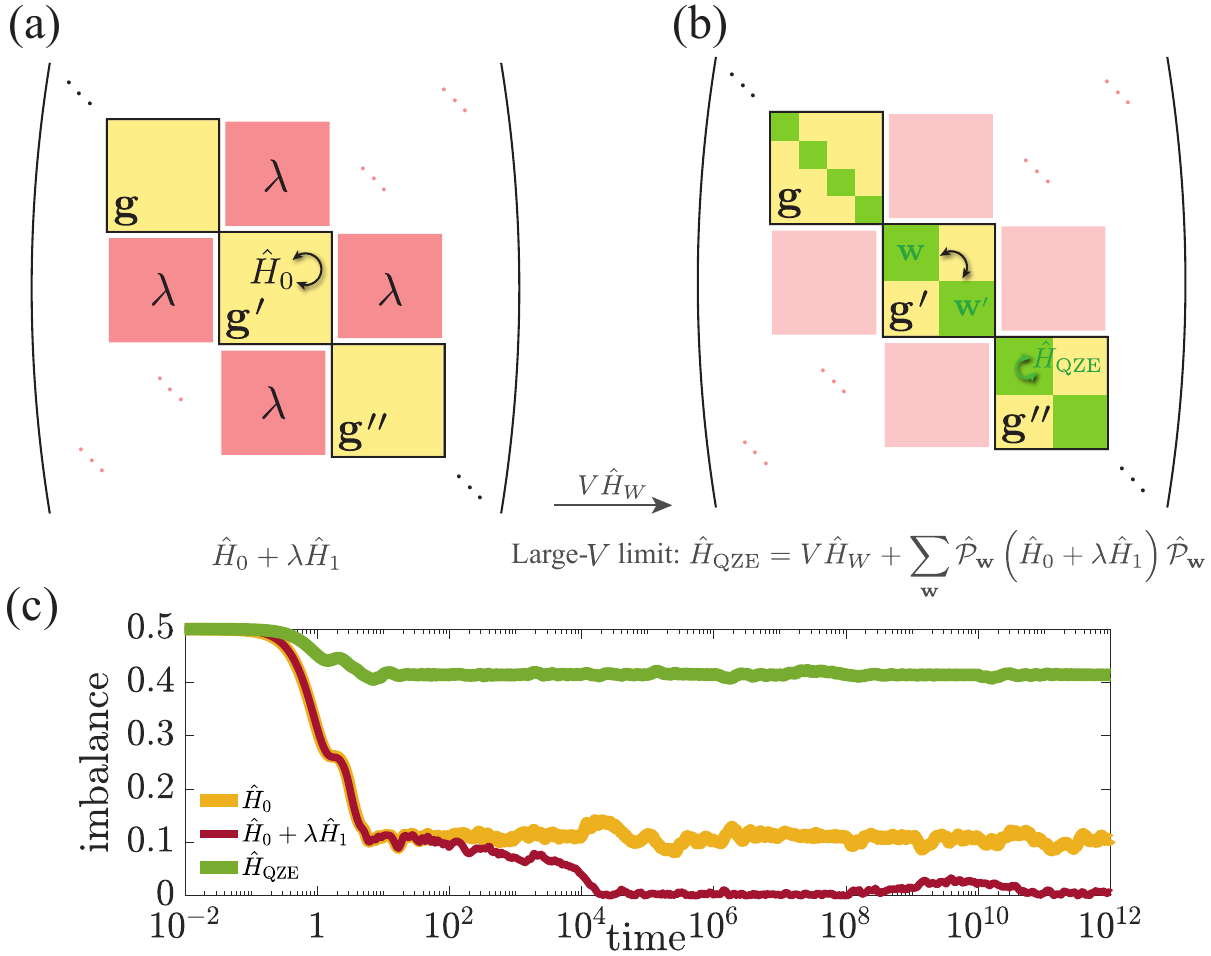}
	\caption{(Color online). The concept of local-pseudogenerator protection of disorder-free localization and its enhancement through dynamically emergent local symmetries. (a) An ideal gauge theory $\hat{H}_0$ of $L$ matter fields and $L$ gauge fields hosts a gauge symmetry with local generator $\hat{G}_j$ whose eigenvalues are $g_j$, a set of which over all local constraints defines a superselection sector $\mathbf{g}=(g_1,g_2,\ldots,g_L)$ with projector $\hat{\mathcal{P}}_\mathbf{g}$ (yellow blocks). Gauge invariance is encoded in the commutation relation $\big[\hat{H}_0,\hat{G}_j\big]=0,\,\forall j$. Disorder-free localization emerges in the quench dynamics of an initial state prepared in a superposition of superselection sectors. A typical quantum synthetic matter realization of $\hat{H}_0$ will include gauge-breaking errors $\lambda\hat{H}_1$ of strength $\lambda$, that will couple the different superselection sectors $\mathbf{g}$, and undermine localized behavior (off-diagonal red blocks). (b) The local pseudogenerator $\hat{W}_j$ acts identically to $\hat{G}_j$ in a given superselection sector (congruent yellow and green blocks), but not necessarily outside of it. The addition of the term $V\hat{H}_W=V\sum_jc_j\hat{W}_j$ of strength $V$, with $c_j$ a properly chosen sequence (see main text), will suppress transitions between different superselection sectors $\mathbf{g}$, thereby protecting localization due to the quantum Zeno effect. At sufficiently large $V$, errors are suppressed and the dynamics is effectively propagated by the emergent gauge theory $\hat{H}_\mathrm{QZE}=V\hat{H}_W+\sum_\mathbf{w}\hat{\mathcal{P}}_\mathbf{w}\big(\hat{H}_0+\lambda\hat{H}_1\big)\hat{\mathcal{P}}_\mathbf{w}$, which has an enhanced local symmetry associated with $\hat{W}_j$, and contains the original $\mathbb{Z}_2$ gauge symmetry generated by $\hat{G}_j$. The eigenvalues $w_j$ of $\hat{W}_j$ define emergent superselection sectors $\mathbf{w}=(w_1,w_2,\ldots,w_L)$ with projectors $\hat{\mathcal{P}}_\mathbf{w}$ (green blocks). The initial state is a superposition of the superselection sectors $\mathbf{w}$, leading to a greater effective disorder over superselection sectors, thereby enhancing disorder-free localization under $\hat{H}_\mathrm{QZE}$. (c) Starting in a translation-invariant bosonic domain-wall state that is a superposition over all gauge superselection sectors, the dynamics of the imbalance under $\hat{H}_0$ will retain memory of the initial state (yellow curve). Errors $\lambda\hat{H}_1$ cause the system to thermalize (red curve). Adding $V\hat{H}_W$ in the large-$V$ limit dynamically induces the emergent theory $\hat{H}_\mathrm{QZE}$ with an enriched local symmetry, thereby enhancing disorder-free localization (green curve, see Sec.~\ref{sec:quench}).}
	\label{fig:schematic}
\end{figure}

Even without quenched disorder, many-body localized behavior can still arise in interacting nonintegrable models \cite{Smith2017,smith2017absence,Metavitsiadis2017,Smith2018,Brenes2018,Schulz2019,Russomanno2020,Papaefstathiou2020,karpov2021disorder,hart2021logarithmic,Zhu2021,Sous2021}. For example, the quench dynamics of a chain of interacting spinless fermions in the presence of a constant electric field will exhibit Stark-MBL \cite{Schulz2019}, and this has been experimentally demonstrated using superconducting qubits \cite{Guo2020stark,Guo2021}. The concept of disorder-free localization has also been extended to lattice gauge theories, through the quench dynamics of translation-invariant product states that form a superposition of gauge superselection sectors \cite{Smith2017,Brenes2018}. This superposition dynamically induces an effective disorder over the background charges associated with the superselection sectors, leading to a strong violation of ETH and indefinite delay of thermalization. The corresponding quench dynamics is restricted to within each sector, and there is no coupling between different superselection sectors. With quantum synthetic matter realizations of gauge theories, it now becomes a realistic possibility to probe disorder-free localization. However, even though gauge symmetry is postulated to be a fundamental property of nature, in quantum simulator realizations it has to be engineered. Indeed, implementations of gauge theories with dynamical matter and gauge fields will always present a certain degree of gauge-breaking errors \cite{aidelsburger2021cold}. Such errors create transitions between different superselection sectors, which undermines disorder-free localization; see Fig.~\ref{fig:schematic}. In the case of perturbative errors, it becomes a prethermal phase, but regardless of the error strength, the system will thermalize eventually \cite{Smith2018}.

Several methods based on energetic constraints have been proposed to control gauge-breaking errors in the context of quench dynamics starting in a given gauge-invariant superselection sector \cite{Zohar2011,Zohar2012,Banerjee2012,Zohar2013,Hauke2013,Stannigel2014,Kuehn2014,Kuno2015,Yang2016,Kuno2017,Negretti2017,Dutta2017,Barros2019,Lamm2020,Halimeh2020e,Kasper2021nonabelian,Halimeh2021gauge,Halimeh2021stabilizing}. A scheme based on a translation-invariant linear sum in the generators $\hat{G}_j$ \cite{Halimeh2020e}, which employs the concept of quantum Zeno subspaces \cite{facchi2002quantum,facchi2004unification,facchi2009quantum,burgarth2019generalized}, has recently been extended to reliably stabilize disorder-free localization \cite{Halimeh2021stabilizingDFL} in spin-$S$ $\mathrm{U}(1)$ quantum link models \cite{Wiese_review,Chandrasekharan1997,Kasper2017}. The concept works based on the quantum Zeno effect: at sufficiently large protection strength, the linear gauge protection term becomes analogous to an external system continually measuring each superselection sector independently of the others, reliably suppressing inter-sector dynamics up to times at least polynomial in the protection strength (see Sec.~\ref{sec:QZE} for details). Hence, the quantum Zeno effect is not only able to protect a single, e.g., ground state, gauge sector \cite{Halimeh2020e}, but in the case of disorder-free localization an extensive number thereof.

In this work, we generalize this approach to $\mathbb{Z}_2$ lattice gauge theories through the use of \textit{local pseudogenerators} $\hat{W}_j$, recently introduced as a powerful tool for gauge protection in these models \cite{Halimeh2021stabilizing}. The pseudogenerator commutes with the full generator of the gauge theory, $\big[\hat{W}_j,\hat{G}_l\big]=0,\,\forall j,l$, but not with the gauge theory itself, $\big[\hat{W}_j,\hat{H}_0\big]\neq0$, and hence $\hat{W}_j$ is not an actual symmetry generator for $\hat{H}_0$. Remarkably, we find that a translation-invariant alternating sum of these pseudogenerators not only protects disorder-free localization \cite{Halimeh2021stabilizingDFL}, but also enhances it through the dynamical emergence of an effective gauge theory that hosts an enhanced local symmetry containing the original $\mathbb{Z}_2$ gauge symmetry. This enhanced local symmetry is due to the $\mathbb{Z}_2$ pseudogenerator $\hat{W}_j$, whose eigenvalues can now be used to define the emergent superselection sectors; see Fig.~\ref{fig:schematic}. We analytically show how this effective theory emerges dynamically from the concept of quantum Zeno subspaces \cite{facchi2002quantum}. Taking advantage of the experimental feasibility of the $\mathbb{Z}_2$ local pseudogenerator, we further propose a readily realizable scheme for the detection of disorder-free localization within accessible timescales of modern Rydberg and superconducting qubit setups.

The rest of the paper is organized as follows: In Sec.~\ref{sec:model}, we introduce the $\mathbb{Z}_2$ lattice gauge theory, which will be the focus of our analysis, along with its local generator and pseudogenerator. In Sec.~\ref{sec:quench}, we present our exact diagonalization results on the quench dynamics of local observables, superselection-sector projectors, and the mid-chain entanglement entropy; we establish the enhancement of disorder-free localization here. In Sec.~\ref{sec:QZE}, we provide the analytic framework of the quantum Zeno subspaces from which we derive the emergent gauge theory. We discuss features of the local-pseudogenerator protection scheme in Sec.~\ref{sec:discussion}. In Sec.~\ref{sec:exp}, we present an experimental proposal for the detection of disorder-free localization in a Rydberg setup based on our scheme. We conclude and provide future outlook in Sec.~\ref{sec:conc}. In addition, we supplement our main conclusions with supporting exact diagonalization results for different errors and larger system sizes in Appendix~\ref{app:supp}, discuss the thermal ensembles we have used in Appendix~\ref{app:thermal}, and provide a discussion of a linear protection scheme based on the local generator in Appendix~\ref{app:full}.

\section{Model and (pseudo)generators}\label{sec:model}
Drawing inspiration from recent experimental \cite{Schweizer2019} and theoretical \cite{Barbiero2019,homeier2020mathbbz2} work, we consider the $(1+1)-$dimensional $\mathbb{Z}_2$ lattice gauge theory given by the Hamiltonian \cite{Zohar2017,Borla2019,Yang2020fragmentation,kebric2021confinement}
\begin{align}\label{eq:H0}
	\hat{H}_0=\sum_{j=1}^{L}\big[J\big(\hat{a}^\dagger_j\hat{\tau}^z_{j,j+1}\hat{a}_{j+1}+\text{H.c.}\big)-h\hat{\tau}^x_{j,j+1}\big],
\end{align}
where $\hat{a}_j^{(\dagger)}$ is the hard-core bosonic annihilation (creation) operator on matter site $j$, and $\hat{n}_j=\hat{a}_j^\dagger \hat{a}_j$ is the corresponding number operator. The electric (gauge) field on the link between matter sites $j$ and $j+1$ is represented by the Pauli matrix $\hat{\tau}_{j,j+1}^{x(z)}$. The generator of the $\mathbb{Z}_2$ gauge symmetry is
\begin{align}\label{eq:Gj}
	\hat{G}_j=(-1)^{\hat{n}_j}\hat{\tau}^x_{j-1,j}\hat{\tau}^x_{j,j+1},
\end{align}
with eigenvalues $g_j=\pm1$, and it defines a local constraint over the matter site $j$ and its adjacent links. The gauge invariance of $\hat{H}_0$ is encoded in the commutation relation $\big[\hat{H}_0,\hat{G}_j\big]=0,\,\forall j$. A set of eigenvalues $\mathbf{g}=(g_1,g_2,\ldots,g_L)$ across local constraints in a gauge theory with $L$ matter sites defines a superselection sector with projector $\hat{\mathcal{P}}_\mathbf{g}$. Disorder-free localization is known to occur in this model when quenching an initial state in a superposition of superselection sectors \cite{Smith2018}.

In typical quantum synthetic matter implementations of gauge theories with dynamical matter and gauge fields, gauge-breaking errors are in practice unavoidable, and these undermine disorder-free localization. For the system we consider, experimentally relevant gauge-breaking terms are of the form
\begin{align}\nonumber
	\lambda\hat{H}_1=\,\lambda\sum_{j=1}^{L}\Big\{&\Big[\hat{a}_j^\dagger\hat{a}_{j+1}\big(\eta_1\hat{\tau}^+_{j,j+1} +\eta_2\hat{\tau}^-_{j,j+1}+1\big)+\mathrm{H.c.}\Big]\\\label{eq:H1}
	&+\big(\eta_3\hat{n}_j-\eta_4\hat{n}_{j+1}+1\big)\hat{\tau}^z_{j,j+1}\Big\},
\end{align}
where $\big[\hat{H}_0,\hat{H}_1\big]\neq0$ and $\big[\hat{G}_j,\hat{H}_1\big]\neq0,\,\forall j$. The coefficients $\eta_{1\ldots4}$ are real numbers that depend on a dimensionless driving parameter $\chi$ employed in the Floquet setup of Ref.~\cite{Schweizer2019}, and they are normalized such that they sum to $1$. Unless otherwise specified, for the results presented in this work we have chosen an experimentally friendly value $\chi=1.84$ set in the experiment of Ref.~\cite{Schweizer2019}, which yields $\eta_1=0.5110$, $\eta_2 = -0.4953$, $\eta_3 = 0.7696$, and $\eta_4 = 0.2147$. However, we have checked that other generic values of these coefficients do not alter the conclusions of our work.

\begin{figure}[t!]
	\centering
	\includegraphics[width=.48\textwidth]{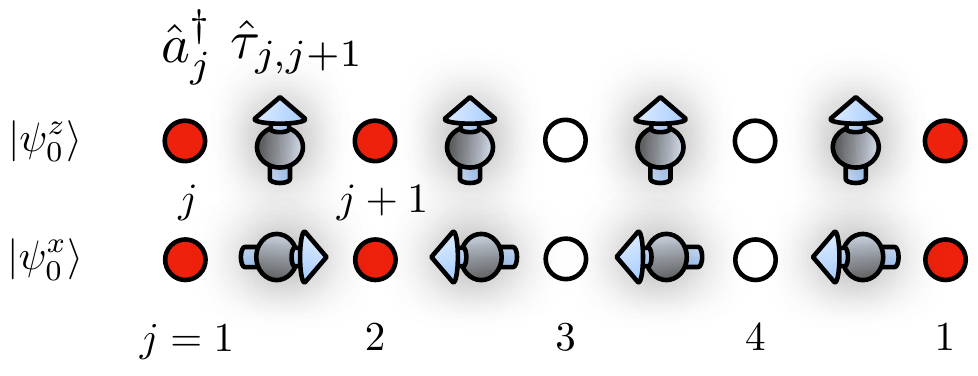}
	\caption{(Color online). The two initial states considered in this work. Both are domain walls on a periodic chain from the perspective of hard-core bosons, with matter sites on the left half of the chain singly occupied (red-filled circles), while those on the right half are empty (blank circles). The main difference lies in the orientation of the electric field on the links between matter sites. In the case of $\ket{\psi^x_0}$, the electric fields are oriented along the positive or negative $x$-direction such that $\hat{G}_j\ket{\psi^x_0}=\ket{\psi^x_0},\,\forall j$, making this initial state gauge-invariant. In contrast, the electric fields are aligned along the positive $z$-direction in the case of $\ket{\psi^z_0}$, rendering the latter a translation-invariant gauge-noninvariant superposition of gauge-invariant superselection sectors (see Table~\ref{Table_G_psi}). Note that both these initial states are product states, and are thus relatively easily implementable in quantum synthetic matter setups.}
	\label{fig:InitialStates} 
\end{figure}

A state $\ket{\psi}$ is said to be in a chosen target sector denoted by $g_j^\text{tar}$ iff $\hat{G}_j\ket{\psi}=g_j^\text{tar}\ket{\psi},\,\forall j$. If we initialize our system in such a state, and probe the ensuing quench dynamics under the \textit{faulty} gauge theory $\hat{H}=\hat{H_0}+\lambda\hat{H}_1+V\hat{H}_W$, we will find gauge violations suppressed $\sim\lambda^2/V^2$ for sufficiently large $V$ using the local-pseudogenerator (LPG) gauge protection
\begin{subequations}
\begin{align}\label{eq:LPGprotection}
\hat{H}_W&=\sum_jc_j\hat{W}_j\\\label{eq:LPG}
\hat{W}_j&=\hat{\tau}^x_{j-1,j}\hat{\tau}^x_{j,j+1}+2g_j^\text{tar}\hat{n}_j,
 \end{align}
\end{subequations}
where $c_j$ is a sequence of appropriately chosen coefficients \cite{Halimeh2021stabilizing}. The LPG $\hat{W}_j$ acts identically to the full generator $\hat{G}_j$ in the target sector, but is experimentally much easier to implement, as it consists of at most two-body single-species terms in contrast to $\hat{G}_j$, which hosts three-body two-species terms. It is important to emphasize that $\hat{W}_j$ is not an actual gauge-symmetry generator of $\hat{H}_0$, and it is easy to check that $\big[\hat{H}_0,\hat{W}_j\big]\neq0$.

It is interesting to note that the $\mathbb{Z}_2$ generator $\hat{G}_j$ and the LPG $\hat{W}_j$ are related according to
\begin{align}
\hat{\mathds{1}}+2g_j^\text{tar}\hat{W}_j-\hat{W}_j^2=2g_j^\text{tar}\hat{G}_j.
\end{align}
As such, it is clear that any Hamiltonian $\hat{H}'$ that commutes with $\hat{W}_j$ must necessarily commute with $\hat{G}_j$,
\begin{align}\label{eq:relation}
    \big[\hat{H}',\hat{W}_j\big]=0,\,\forall j\implies \big[\hat{H}',\hat{G}_j\big]=0,\,\forall j.
\end{align}
This relation states that if the Hamiltonian $\hat{H}'$ hosts a local symmetry associated with $\hat{W}_j$, then it will surely host the $\mathbb{Z}_2$ gauge symmetry generated by $\hat{G}_j$. In other words, the local symmetry associated with $\hat{W}_j$ contains the $\mathbb{Z}_2$ gauge symmetry generated by $\hat{G}_j$. Note that the converse of relation~\eqref{eq:relation} is not necessarily true, and $\hat{H}_0$ is one counterexample.

For the rest of our discussion and without loss of generality, we will set $g_j^\text{tar}=1$ in Eq.~\eqref{eq:LPG}---we have checked that our conclusions remain the same for other values of it. Moreover, we will use the sequence $c_j=[6(-1)^j+5]/11$ \cite{Halimeh2021stabilizing} unless otherwise specified.

\section{Quench dynamics}\label{sec:quench}
We will now demonstrate how LPG protection can be employed to not only stabilize disorder-free localization, but also to enhance it through a dynamically emergent enhanced local symmetry associated with the LPG and containing the original $\mathbb{Z}_2$ gauge symmetry.

\subsection{Imbalance}
To probe disorder-free localization, we calculate in exact diagonalization the quench dynamics, as propagated by the faulty theory $\hat{H}$, of the spatiotemporally averaged imbalance in the boson number between both halves of the chain, defined as
\begin{align}\label{eq:imbalance}
\mathcal{I}(t)=\frac{1}{Lt}\int_0^t ds\sum_{j=1}^Lp_j\bra{\psi(s)}\hat{n}_j\ket{\psi(s)},
\end{align}
where $p_j=2\bra{\psi_0}\hat{n}_j\ket{\psi_0}-1$, $\ket{\psi_0}=\ket{\psi_0^{x,z}}$ is one of the two initial states we consider in this work (see Fig.~\ref{fig:InitialStates}), $L$ is the number of matter sites (and also the number of gauge links as we employ periodic boundary conditions), and $\ket{\psi(t)}=e^{-i\hat{H}t}\ket{\psi_0}$. For the following results, we will set $L=4$, while relegating results for larger system sizes to Appendix~\ref{app:supp}.

\begin{table}[t!]
	\centering
	\begin{tabular}{|| c || c | c ||}
		\hline
		 $\mathbf{g}=(g_1,g_2,g_3,g_4)$ & $\bra{\psi^x_0}\hat{\mathcal{P}}_\mathbf{g}\ket{\psi^x_0}$ & $\bra{\psi^z_0}\hat{\mathcal{P}}_\mathbf{g}\ket{\psi^z_0}$ \\ [0.5ex] 
		\hline\hline
		$(-1,-1,-1,-1)$ & $0$ & $0.125$\\ 
		\hline
		$(-1,-1,+1,+1)$ & $0$ & $0.125$\\ 
		\hline
		$(-1,+1,-1,+1)$ & $0$ & $0.125$\\ 
		\hline
		$(-1,+1,+1,-1)$ & $0$ & $0.125$\\ 
		\hline
		$(+1,-1,-1,+1)$ & $0$ & $0.125$\\ 
		\hline
		$(+1,-1,+1,-1)$ & $0$ & $0.125$\\ 
		\hline
		$(+1,+1,-1,-1)$ & $0$ & $0.125$\\ 
		\hline
		$(+1,+1,+1,+1)$ & $1$ & $0.125$\\  [1ex]
		\hline
	\end{tabular}
	\caption{The accessible superselection sectors $\mathbf{g}$ and the expectation values of their projectors $\hat{\mathcal{P}}_\mathbf{g}$ relative to the initial states of Fig.~\ref{fig:InitialStates}. The gauge-invariant initial state $\ket{\psi^x_0}$ resides within a single superselection sector $g_j=+1,\,\forall j$, while the gauge-noninvariant initial state $\ket{\psi^z_0}$ is an equal-weight superposition over all accessible superselection sectors.}
	\label{Table_G_psi}
\end{table}

\begin{figure}[t!]
	\centering
	\includegraphics[width=.48\textwidth]{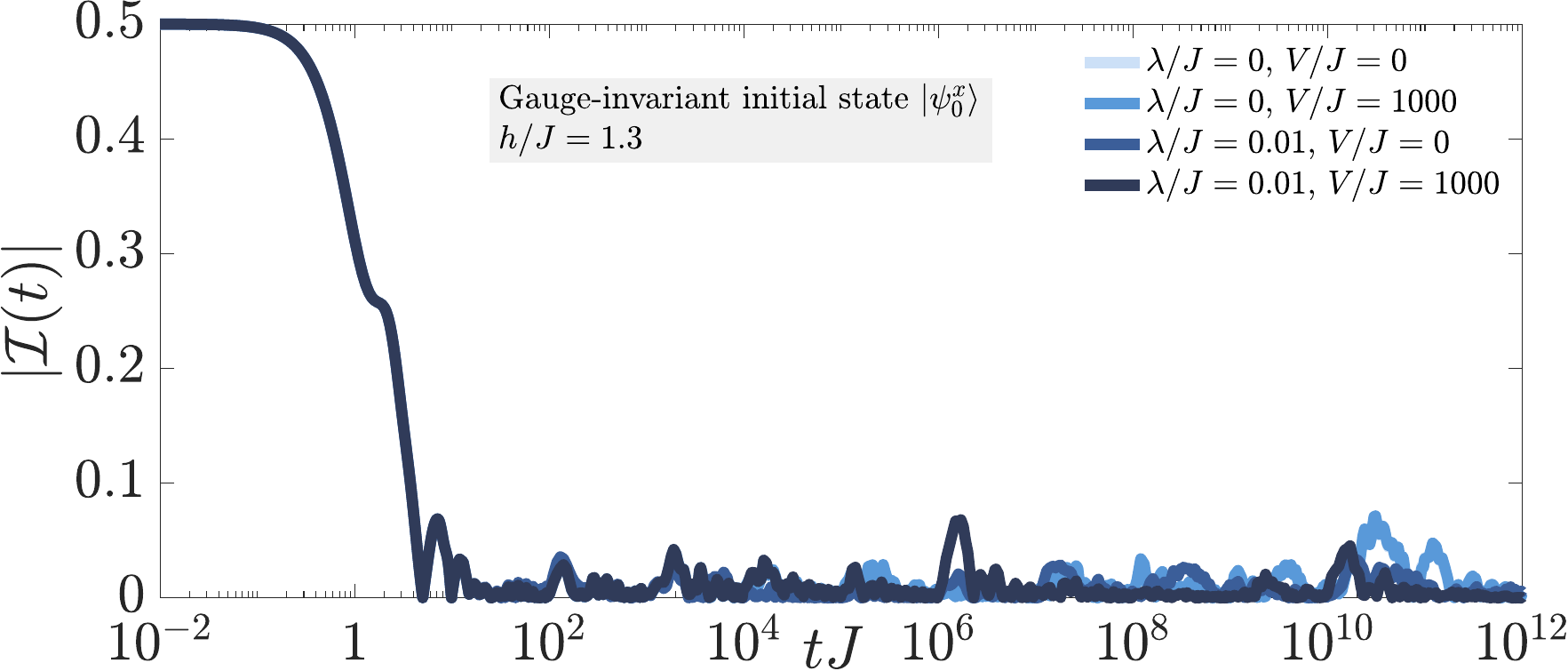}
	\caption{(Color online). Quench dynamics under $\hat{H}=\hat{H}_0+\lambda\hat{H}_1+V\hat{H}_W$ of the imbalance~\eqref{eq:imbalance} starting in the gauge-invariant domain-wall state $\ket{\psi^x_0}$, see Fig.~\ref{fig:InitialStates}. As predicted by the corresponding thermal ensembles, the system thermalizes with the imbalance vanishing regardless of what values $\lambda$ and $V$ take.}
	\label{fig:GI_imbalance} 
\end{figure}

Let us first prepare our system in the initial state $\ket{\psi^x_0}$ in the superselection sector $\hat{G}_j\ket{\psi^x_0}=\ket{\psi^x_0},\,\forall j$, such that it is a domain-wall state at half-filling from the perspective of hard-core bosons; see Fig.~\ref{fig:InitialStates} and Table~\ref{Table_G_psi}. In the wake of a quench with $\hat{H}=\hat{H}_0+\lambda\hat{H}_1+V\hat{H}_W$, the system is expected to thermalize, and indeed we find that the imbalance relaxes to zero as predicted by the corresponding thermal ensembles, see Fig.~\ref{fig:GI_imbalance} and Appendix~\ref{app:thermal}. 

\begin{figure*}[t!]
	\centering
	\includegraphics[width=.48\textwidth]{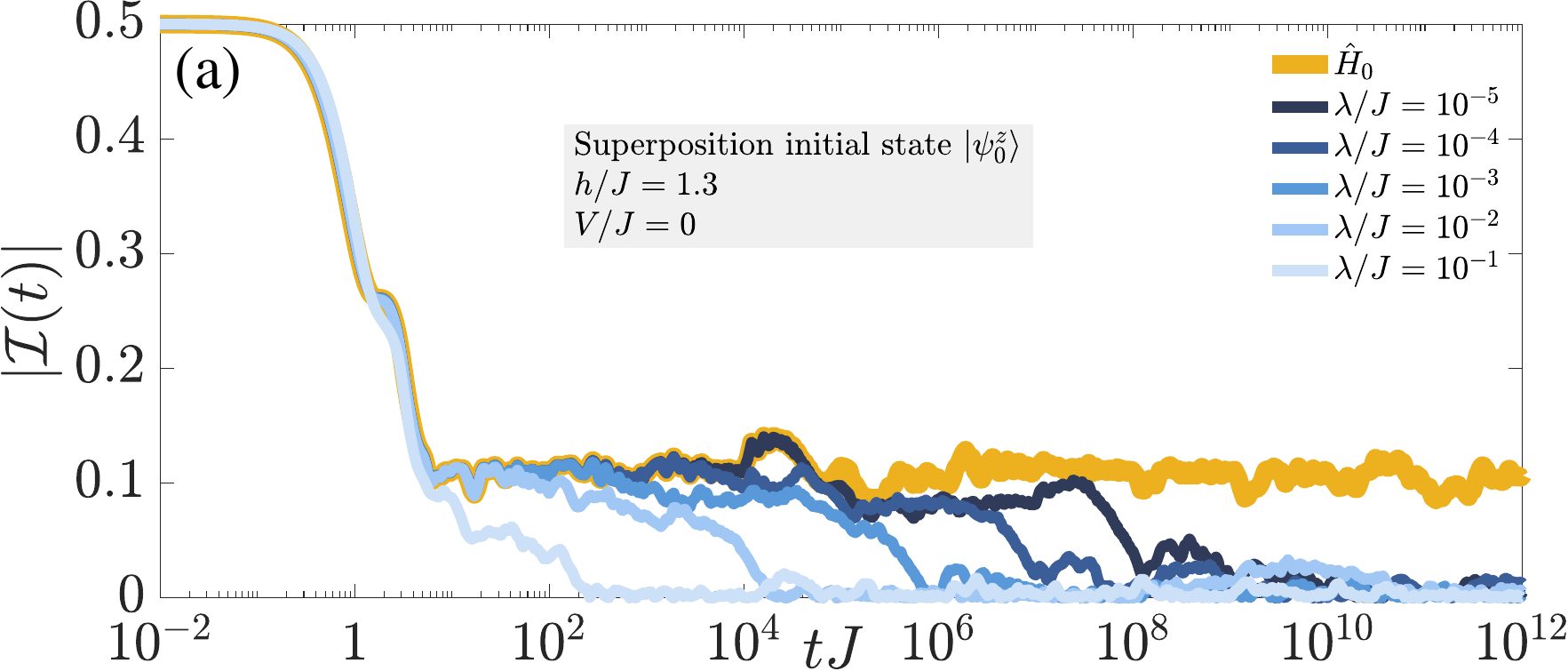}\quad
	\includegraphics[width=.48\textwidth]{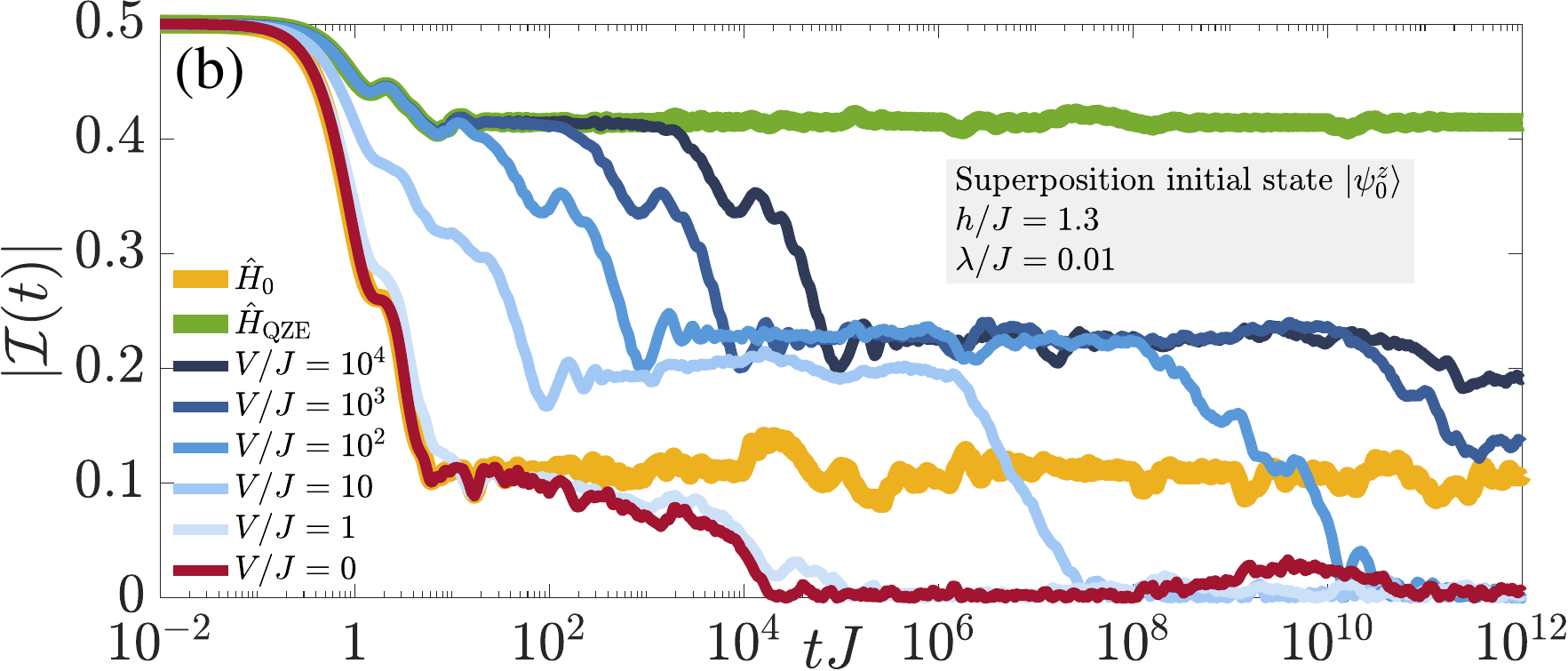}\\
	\vspace{0.1cm}
	\includegraphics[width=.48\textwidth]{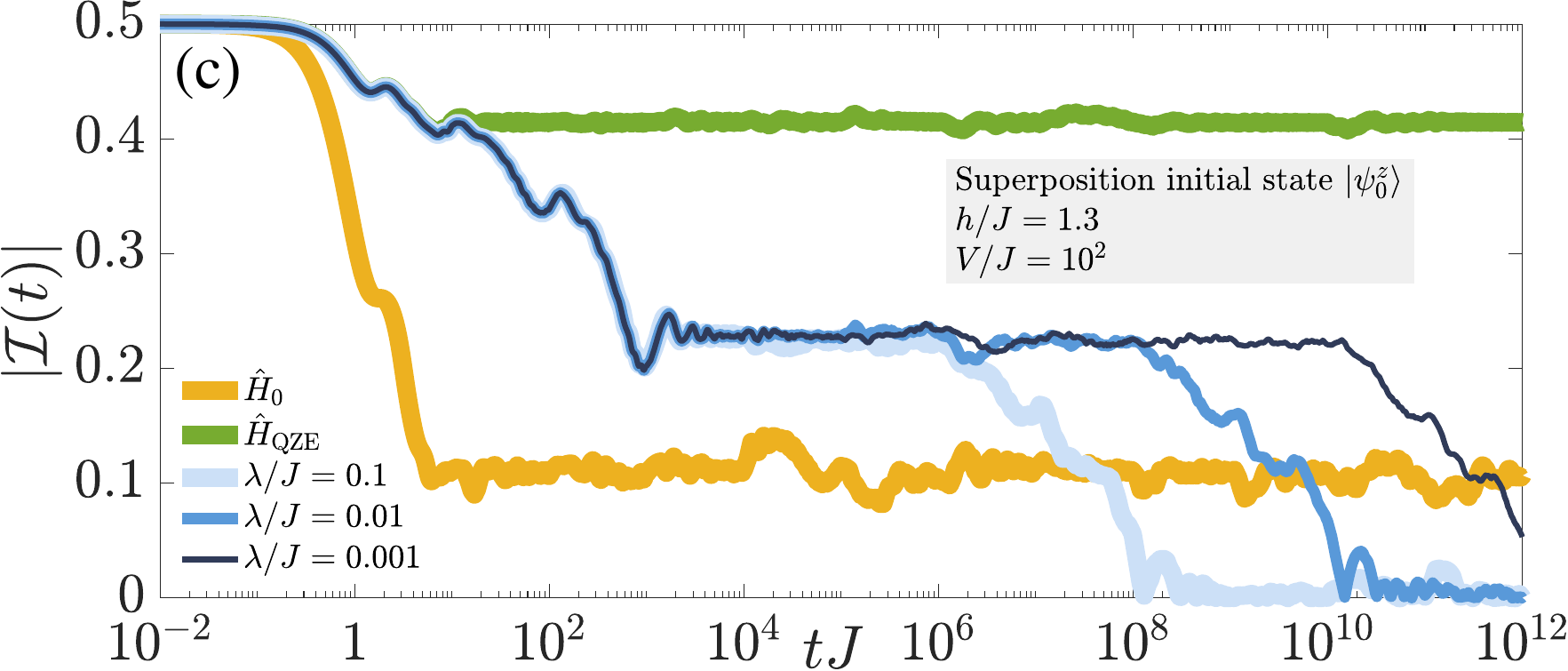}\quad
	\includegraphics[width=.48\textwidth]{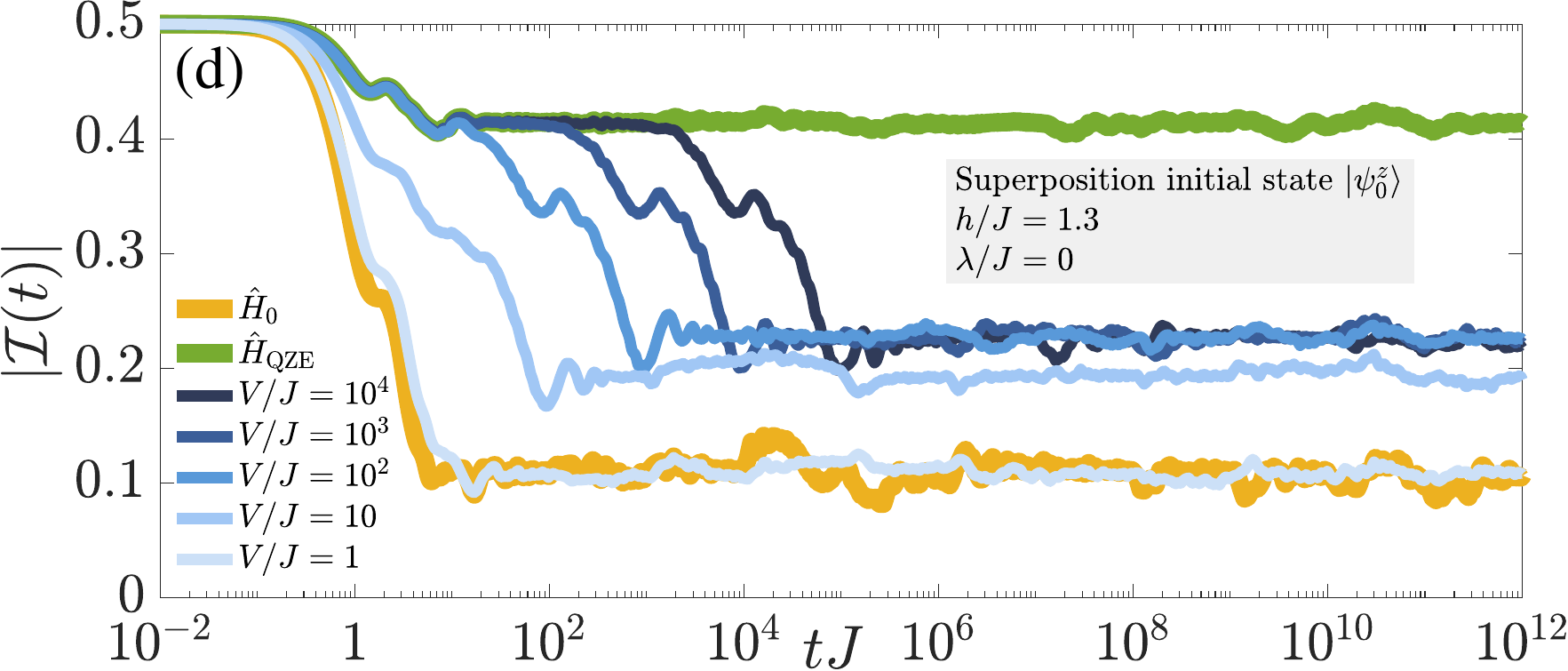}
	\caption{(Color online). Quench dynamics under the faulty theory $\hat{H}=\hat{H}_0+\lambda\hat{H}_1+V\hat{H}_W$ of the imbalance~\eqref{eq:imbalance} starting in $\ket{\psi^z_0}$, a bosonic domain-wall state where the electric flux on each link is a superposition of both its possible configurations, rendering $\ket{\psi^z_0}$ a superposition over all physical superselection sectors. (a) This state leads to disorder-free localization up to all accessible times when there are no gauge-breaking errors. In the presence of the latter, disorder-free localization exhibits staircase prethermalization, and the system eventually thermalizes with a vanishing imbalance. (b,c,d) Introducing LPG protection prolongs the prethermal disorder-free localization plateau, and even enhances it. We find that initially an emergent gauge theory $\hat{H}_\mathrm{QZE}=\sum_\mathbf{w}\hat{\mathcal{P}}_\mathbf{w}\big(\hat{H}_0+\lambda\hat{H}_1\big)\hat{\mathcal{P}}_\mathbf{w}$, with an enhanced local symmetry associated with $\hat{W}_j$ and containing the original $\mathbb{Z}_2$ gauge symmetry, effectively describes the dynamics up to a timescale $\propto V/J^2$, after which another effective $\mathbb{Z}_2$ gauge theory emerges lasting (b,c) up to a timescale $\propto V^2/(\lambda^2 J)$ in the presence of gauge-breaking errors, and (d) indefinitely when there are no errors.}
	\label{fig:nonGI_imbalance} 
\end{figure*}

Let us now investigate what happens when the system is initialized in $\ket{\psi^z_0}$, which is also a domain-wall state in the hard-core bosons, but its electric fields all point in the positive $z$-direction, creating an equal-weight superposition over all physical superselection sectors corresponding to $\hat{G}_j$; see Fig.~\ref{fig:InitialStates} and Table~\ref{Table_G_psi}. Such an initial state is known to lead to disorder-free localization \cite{Smith2018}. Indeed, we see in Fig.~\ref{fig:nonGI_imbalance} that in the absence of gauge-breaking errors, the system retains memory of the initial state up to all accessible evolution times (yellow curve), with the imbalance relaxing to a value of approximately $0.1$, despite the corresponding thermal ensembles predicting a vanishing imbalance. However, once gauge-breaking errors are present, disorder-free localization becomes a prethermal phase as shown in Fig.~\ref{fig:nonGI_imbalance}(a), reminiscent of staircase prethermalization, which has previously been observed in gauge theories in a different setting \cite{Halimeh2020b,Halimeh2020c,Halimeh2021stabilizingDFL}. In particular, we find that at a given value of the error strength $\lambda$, the imbalance leaves the error-free plateau at a timescale $\propto1/\lambda$, after which it enters a second prethermal plateau and finally decays to zero, as predicted by the corresponding thermal ensembles, at a timescale $\propto J/\lambda^2$.

Fixing the error strength to $\lambda=0.01J$, we now study the efficacy of the LPG gauge protection \eqref{eq:LPGprotection} in stabilizing disorder-free localization for a quench by $\hat{H}$ starting in the initial state $\ket{\psi^z_0}$. Remarkably, as shown in Fig.~\ref{fig:nonGI_imbalance}(b), not only is the disorder-free localization prolonged with larger $V$, it is also enhanced, with the prethermal disorder-free localization plateaus exhibiting greater memory of the initial state by taking on larger values than the ideal case (yellow curve) before thermalization.

Specifically, we find that up to a timescale $\propto V/J^2$, the dynamics of the imbalance is effectively described by the emergent gauge theory $\hat{H}_\mathrm{QZE}=\sum_\mathbf{w}\hat{\mathcal{P}}_\mathbf{w}\big(\hat{H}_0+\lambda\hat{H}_1\big)\hat{\mathcal{P}}_\mathbf{w}$, which hosts an enhanced local symmetry associated with the LPG $\hat{W}_j$. As a result of relation~\eqref{eq:relation}, this local symmetry contains the $\mathbb{Z}_2$ gauge symmetry generated by $\hat{G}_j$. Indeed, $\hat{H}_\mathrm{QZE}$ can be derived through the quantum Zeno effect \cite{facchi2002quantum,facchi2004unification,facchi2009quantum,burgarth2019generalized}, as we demonstrate in Sec.~\ref{sec:QZE}, and satisfies $\big[\hat{H}_\mathrm{QZE},\hat{G}_j\big]=\big[\hat{H}_\mathrm{QZE},\hat{W}_j\big]=0,\,\forall j$. Here, $\mathbf{w}=(w_1,w_2,\ldots,w_L)$ are superselection sectors of $\hat{W}_j$ with projectors $\hat{\mathcal{P}}_\mathbf{w}$. We see that the timescale of this emergent theory does not seem to depend on $\lambda$, and this is because the dominant gauge-breaking term for the emergent local symmetry is $\hat{H}_0$ (see Sec.~\ref{sec:QZE} for further details). After this timescale, another effective gauge theory emerges with only the $\mathbb{Z}_2$ gauge symmetry, that persists up to a timescale $\propto V^2/(\lambda^2 J)$, as can be discerned from Fig.~\ref{fig:nonGI_imbalance}(b,c), after which thermalization occurs and the imbalance vanishes.

In the absence of gauge-breaking errors, we observe in Fig.~\ref{fig:nonGI_imbalance}(d) the same behavior up to a timescale $\propto V/J^2$, where $\hat{H}_\mathrm{QZE}=\sum_\mathbf{w}\hat{\mathcal{P}}_\mathbf{w}\hat{H}_0\hat{\mathcal{P}}_\mathbf{w}$ faithfully reproduces the dynamics as analytically predicted in Sec.~\ref{sec:QZE}. However, after this timescale once again an emergent $\mathbb{Z}_2$ gauge theory arises that lasts indefinitely. We note that we have checked that the thermal ensembles corresponding to a quench of $\ket{\psi^z_0}$ by $\hat{H}_0+V\hat{H}_W$ predict a vanishing imbalance, and therefore the nonzero-imbalance plateaus in Fig.~\ref{fig:nonGI_imbalance}(d) cannot be thermal, but rather are a signature of localized dynamics.

This leads to the remarkable conclusion that employing LPG protection not only preserves the original gauge symmetry of the ideal theory up to a timescale $\propto V^2/(\lambda^2 J)$, thereby protecting disorder-free localization, but it also dynamically induces an emergent local symmetry up to a timescale $\propto V/J^2$ that further enhances disorder-free localization. This is possible because the initial state can be viewed as a superposition over the superselection sectors of $\hat{W}_j$ as well as those of $\hat{G}_j$ (see Table~\ref{Table_W_psi}). When the dynamics is propagated by only $\hat{H}_0$, only the superposition over the superselection sectors $\mathbf{g}$ will induce an effective disorder since $\big[\hat{H}_0,\hat{G}_j\big]=0,\,\forall j$, but $\big[\hat{H}_0,\hat{W}_j\big]\neq0$. On the other hand, once the dynamics is propagated by the faulty theory $\hat{H}=\hat{H}_0+\lambda\hat{H}_1+V\hat{H}_W$ at sufficiently large $V$, one can derive through the quantum Zeno effect that the dynamics is effectively reproduced by
\begin{align}\label{eq:effective}
\hat{H}_\mathrm{QZE}=V\hat{H}_W+\sum_\mathbf{w}\hat{\mathcal{P}}_\mathbf{w}\big(\hat{H}_0+\lambda\hat{H}_1\big)\hat{\mathcal{P}}_\mathbf{w},
\end{align}
with an error of upper bound $\propto t V_0^2L^2/V$ \cite{Halimeh2021stabilizing}, with $V_0$ an energetic term that is roughly a linear sum in $J$, $\lambda$, and $h$; cf.~Sec.~\ref{sec:QZE}. This emergent gauge theory enhances disorder-free localization through its enhanced local symmetry associated with $\hat{W}_j$, because now the superposition over the superselection sectors $\mathbf{w}$ will induce a further effective disorder in addition to that due to the superposition over the superselection sectors $\mathbf{g}$. We note that our exact diagonalization results show that the term $V\hat{H}_W$ is inconsequential to the dynamics of the local observables under Eq.~\eqref{eq:effective}, which is why we have neglected this term in $\hat{H}_\mathrm{QZE}$ for Fig.~\ref{fig:nonGI_imbalance}. The enhancement of disorder-free localization through LPG protection is a main result of this work, which we will analyze in more detail below.

\begin{table}[t!]
	\centering
	\begin{tabular}{|| c || c | c ||}
		\hline
		 $\mathbf{w}=(w_1,w_2,w_3,w_4)$ & $\bra{\psi^x_0}\hat{\mathcal{P}}_\mathbf{w}\ket{\psi^x_0}$ & $\bra{\psi^z_0}\hat{\mathcal{P}}_\mathbf{w}\ket{\psi^z_0}$ \\ [0.5ex] 
		\hline\hline
		$(+1,+1,-1,-1)$ & $0$ & $0.125$\\ 
		\hline
		$(+1,+1,+1,+1)$ & $1$ & $0.125$\\ 
		\hline
		$(+1,+3,-1,+1)$ & $0$ & $0.125$\\ 
		\hline
		$(+1,+3,+1,-1)$ & $0$ & $0.125$\\ 
		\hline
		$(+3,+1,-1,+1)$ & $0$ & $0.125$\\ 
		\hline
		$(+3,+1,+1,-1)$ & $0$ & $0.125$\\ 
		\hline
		$(+3,+3,-1,-1)$ & $0$ & $0.125$\\ 
		\hline
		$(+3,+3,+1,+1)$ & $0$ & $0.125$\\  [1ex] 
		\hline
	\end{tabular}
	\caption{The superselection sectors $\mathbf{w}$ of the gauge symmetry generated by $\hat{W}_j$, and the expectation values of their projectors relative to the initial states of Fig.~\ref{fig:InitialStates}. The gauge-invariant initial state $\ket{\psi^x_0}$ resides within a single superselection sector $w_j=g_j=+1,\,\forall j$, while the gauge-noninvariant initial state $\ket{\psi^z_0}$ is an equal-weight superposition over various superselection sectors $\mathbf{w}$ that are equivalent to the superselection sectors in Table~\ref{Table_G_psi}.}
	\label{Table_W_psi}
\end{table}

\begin{figure*}[t!]
	\centering
	\includegraphics[width=.48\textwidth]{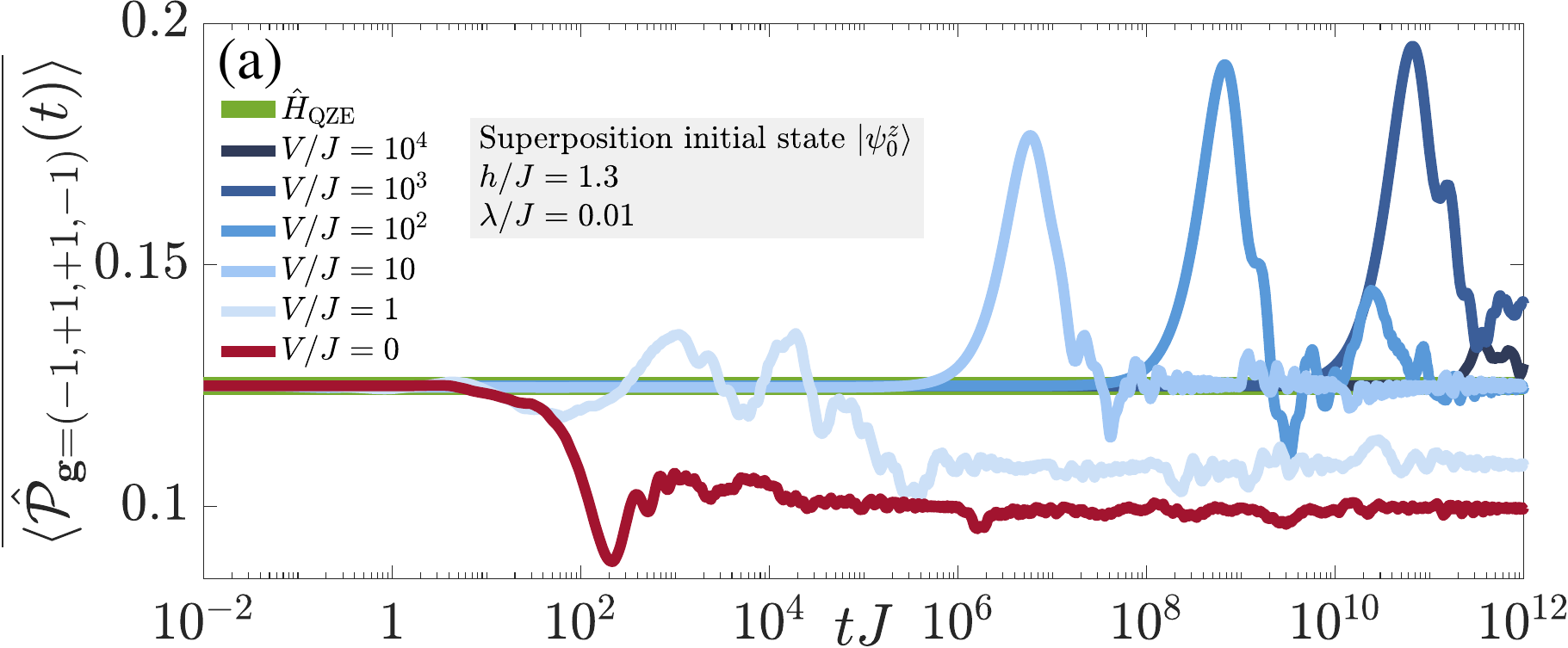}\quad
	\includegraphics[width=.48\textwidth]{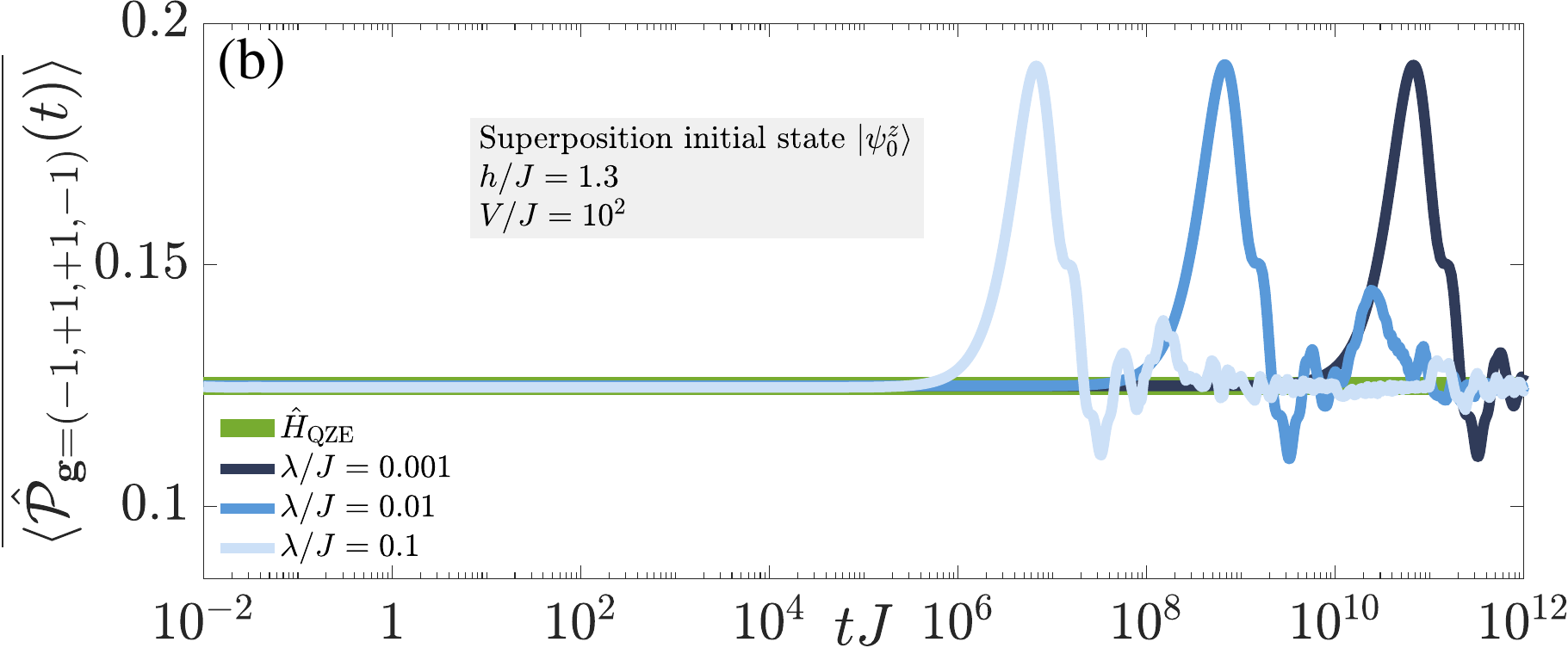}\\
	\vspace{0.1cm}
	\includegraphics[width=.48\textwidth]{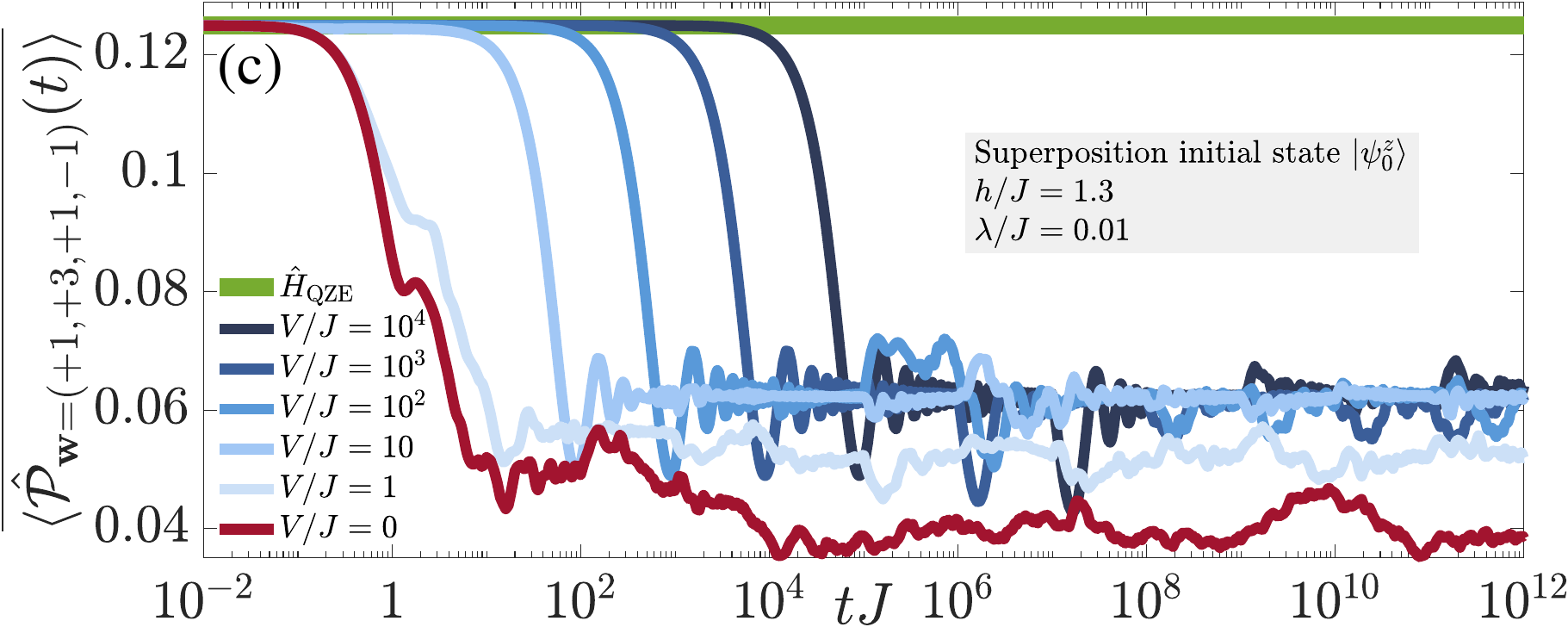}\quad\includegraphics[width=.48\textwidth]{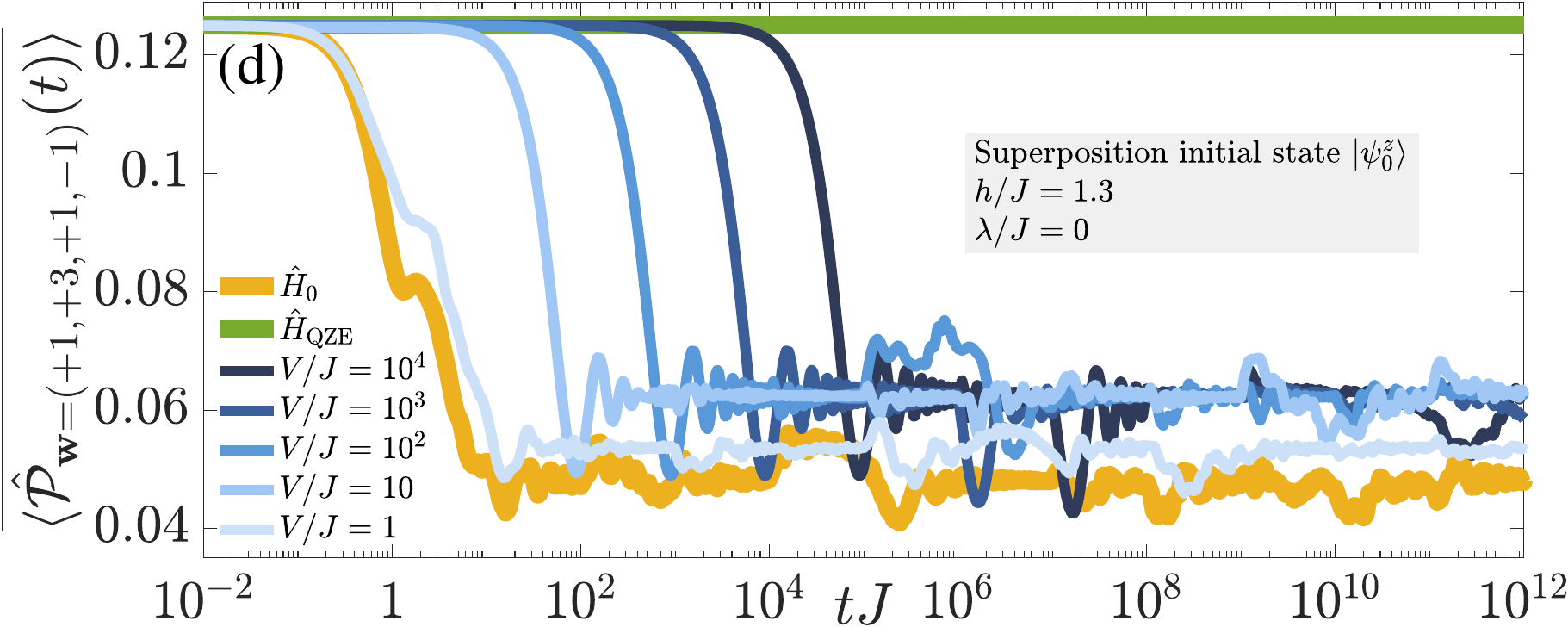}
	\caption{(Color online). Quench dynamics of the projectors onto (a,b) the superselection sectors $\mathbf{g}$ of $\hat{G}_j$ and (c,d) the superselection sectors $\mathbf{w}$ of $\hat{W}_j$. (a,b) The quench dynamics of $\hat{\mathcal{P}}_\mathbf{g}$ under the faulty theory $\hat{H}=\hat{H}_0+\lambda\hat{H}_1+V\hat{H}_W$ restores its original value with larger $V$ up to a timescale $\propto V^2/(\lambda^2J)$, which exceeds the analytic prediction from the quantum Zeno effect (see Sec.~\ref{sec:QZE}). The timescale up to which $\langle\hat{\mathcal{P}}_\mathbf{g}\rangle$ retains its initial value matches that at which the imbalance leaves its second prethermal plateau and begins to thermalize to zero, as shown in Fig.~\ref{fig:nonGI_imbalance}(b). The quench dynamics of $\hat{\mathcal{P}}_\mathbf{w}$ at (c) $\lambda=0.01J$ and (d) $\lambda=0$ indicate a common timescale $\propto V/J^2$ at which $\langle\hat{\mathcal{P}}_\mathbf{w}\rangle$ deviates from its initial value, showing that it depends more strongly on $\hat{H}_0$ than the gauge-breaking term $\lambda\hat{H}_1$. These results confirm that up to a timescale $\propto V/J^2$, an emergent gauge theory $\hat{H}_\mathrm{QZE}$ faithfully reproduces the quench dynamics, where this theory hosts an enhanced local symmetry associated with $\hat{W}_j$ that contains the $\mathbb{Z}_2$ gauge symmetry of $\hat{H}_0$. After this timescale, the enhanced local symmetry reduces to the original $\mathbb{Z}_2$ gauge symmetry, and a new renormalized $\mathbb{Z}_2$ gauge theory emerges.}
	\label{fig:projectors} 
\end{figure*}

\subsection{Superselection-sector projectors}
In order to get a deeper understanding of the behavior and timescales exhibited in Fig.~\ref{fig:nonGI_imbalance}, we study the expectation values of the projectors $\hat{\mathcal{P}}_\mathbf{g}$ and $\hat{\mathcal{P}}_\mathbf{w}$ onto the superselection sectors $\mathbf{g}$ and $\mathbf{w}$, respectively. The corresponding data is shown in Fig.~\ref{fig:projectors}(a,c) for $\mathbf{g}=(-1,+1,+1,-1)$ and $\mathbf{w}=(+1,+3,+1,-1)$, respectively, at a fixed gauge-breaking error $\lambda=0.01J$. Under the effective Hamiltonian $\hat{H}_\mathrm{QZE}$, the expectation values of both $\hat{\mathcal{P}}_\mathbf{g}$ and $\hat{\mathcal{P}}_\mathbf{w}$ will always be equal to their initial values, because $\hat{H}_\mathrm{QZE}$ hosts an enhanced local symmetry associated with $\hat{W}_j$, and this local symmetry contains the $\mathbb{Z}_2$ gauge symmetry generated by $\hat{G}_j$. In other words, $\hat{H}_\mathrm{QZE}$ restricts the dynamics of the total system to intra-sector dynamics, and does not couple different sectors. The dynamics under the faulty theory $\hat{H}=\hat{H}_0+\lambda\hat{H}_1+V\hat{H}_W$ can be shown analytically to be well-reproduced by $\hat{H}_\mathrm{QZE}$ up to timescales linear in $V$ (see Sec.~\ref{sec:QZE}). Therefore, we can expect that $\langle\hat{\mathcal{P}}_\mathbf{g}\rangle$ and $\langle\hat{\mathcal{P}}_\mathbf{w}\rangle$ will remain controllably near their initial values until \textit{at least} such timescales, and indeed this is what we see in Fig.~\ref{fig:projectors}. Interestingly though, we see that $\langle\hat{\mathcal{P}}_\mathbf{w}\rangle$ deviates from its initial value much earlier than $\langle\hat{\mathcal{P}}_\mathbf{g}\rangle$. This is not surprising because enhanced local symmetry associated with $\hat{W}_j$ is subjected not only to the errors $\lambda\hat{H}_1$, but also to the more dominant $\hat{H}_0$, which is an error term from the viewpoint of this local symmetry. As such, the enhanced local symmetry is expected to reduce to the $\mathbb{Z}_2$ gauge symmetry before the latter is also broken by $\lambda\hat{H}_1$ at sufficiently long times.

This revelation also allows us to understand exactly what is happening in the results of Fig.~\ref{fig:nonGI_imbalance}(b,d). The quench dynamics of the imbalance under $\hat{H}$ is initially reproduced by $\hat{H}_\mathrm{QZE}$ up to timescales linear in $V$, but then this plateau decays into one lower in value. This occurs at the same timescale $\langle\hat{\mathcal{P}}_\mathbf{w}\rangle$ leaves its initial value in Fig.~\ref{fig:projectors}(c) at finite $\lambda$, and also in Fig.~\ref{fig:projectors}(d) for the error-free case. As such, we find that at sufficiently large $V$, a dynamically induced effective gauge theory arises up to timescales $\propto V/J^2$ that has an enhanced local symmetry associated with $\hat{W}_j$, and as a consequence, also hosts the $\mathbb{Z}_2$ gauge symmetry generated by $\hat{G}_j$. The latter is subjected to only one gauge-breaking term $\lambda\hat{H}_1$, while the former is undermined by two gauge-breaking terms $\hat{H}_0+\lambda\hat{H}_1$. After a timescale $\propto V/J^2$, the emergent gauge theory is replaced by a renormalized gauge theory with only the original $\mathbb{Z}_2$ gauge symmetry intact, which is further compromised at a timescale $\propto V^2/(\lambda^2 J)$, see Fig.~\ref{fig:projectors}(a,b), as in Fig.~\ref{fig:nonGI_imbalance}(b,c). However, this numerically obtained timescale also depends on the error term (see Appendix.~\ref{app:supp}).

\begin{figure}[t!]
	\centering
	\includegraphics[width=.48\textwidth]{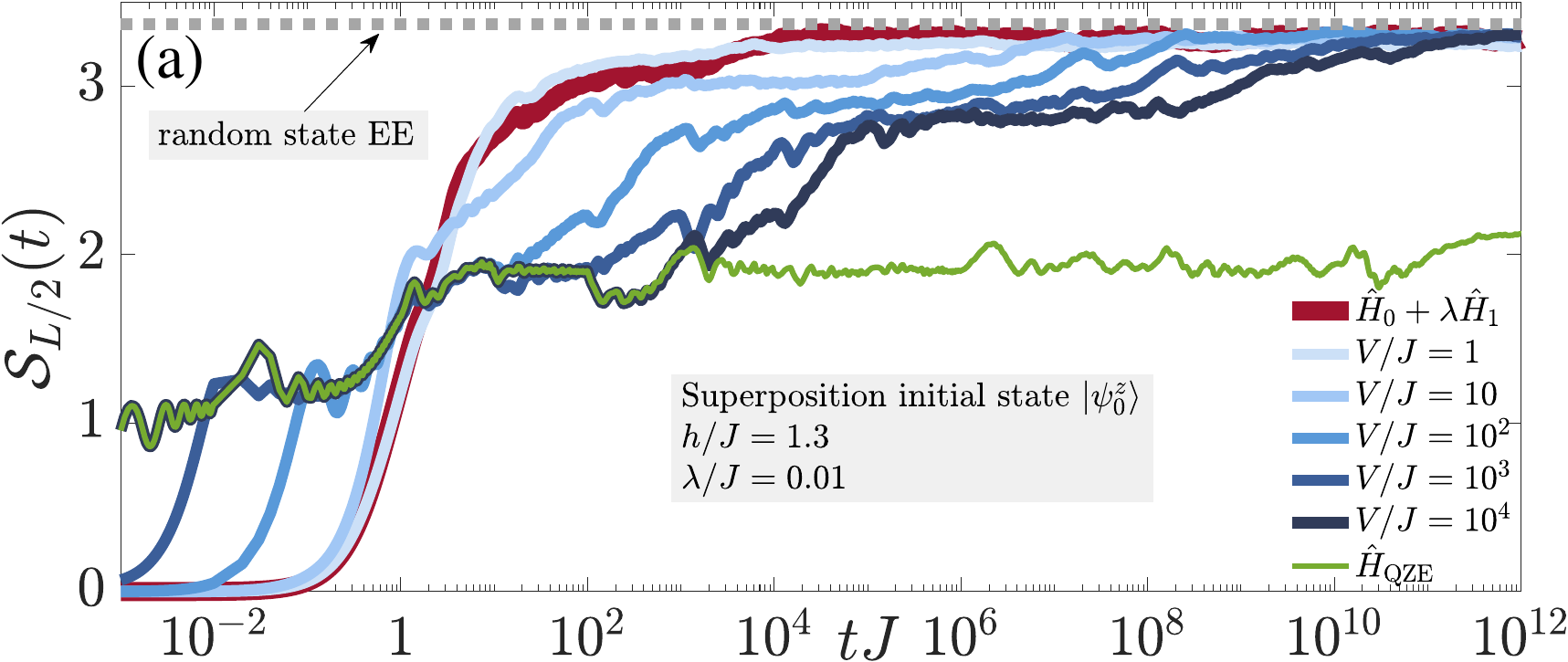}\\
	\vspace{0.1cm}
	\includegraphics[width=.48\textwidth]{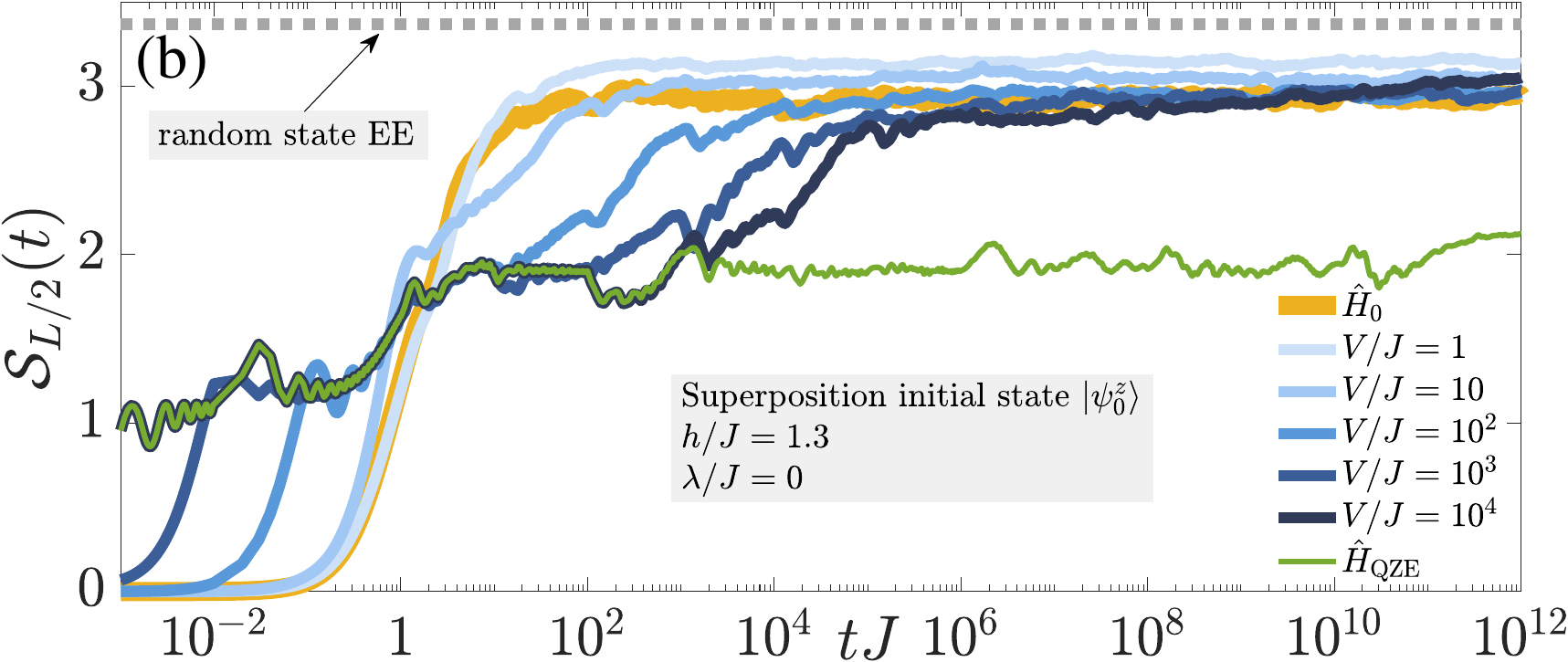}
	\caption{(Color online). Quench dynamics of the mid-chain von Neumann entanglement entropy under $\hat{H}=\hat{H}_0+\lambda\hat{H}_1+V\hat{H}_W$, where $\lambda\hat{H}_1$ is given by Eq.~\eqref{eq:H1}, (a) in the presence of errors at $\lambda=0.01J$ and (b) without errors, $\lambda=0$. (a) Without protection, the error leads to the entanglement entropy growing to its maximal value (red curve) such that it equals the entanglement entropy of a random pure state (dotted gray line). Upon adding LPG protection (shades of blue), we see that with larger $V$ the entanglement entropy takes longer to reach this maximal value. At sufficiently large $V$, the dynamics of the entanglement entropy is well-reproduced by the effective Hamiltonian $\hat{H}_\mathrm{QZE}=V\hat{H}_W+\sum_\mathbf{w}\hat{\mathcal{P}}_\mathbf{w}\big(\hat{H}_0+\lambda\hat{H}_1\big)\hat{\mathcal{P}}_\mathbf{w}$ up to timescales linear in $V$, after which a renormalized $\mathbb{Z}_2$ gauge theory takes over leading to a plateau in the entanglement entropy lasting up to the timescale $\propto V^2/(\lambda^2 J)$, after which the entanglement entropy grows to its maximal value. Under $\hat{H}_\mathrm{QZE}$, the entanglement entropy saturates to a value much smaller than the unprotected case (green curve). (b) In the error-free case, the entanglement entropy without LPG protection reaches a maximal value lower than that of a random pure state due to intact disorder-free localization (yellow curve). Upon adding LPG protection, disorder-free localization is enhanced at sufficiently large $V$, leading to an entanglement entropy well-reproduced up to timescales linear in $V$ by $\hat{H}_\mathrm{QZE}=V\hat{H}_W+\sum_\mathbf{w}\hat{\mathcal{P}}_\mathbf{w}\hat{H}_0\hat{\mathcal{P}}_\mathbf{w}$, under which the entanglement entropy saturates at a much smaller value than the case without LPG protection. After this linear-in-$V$ timescale, the entanglement entropy saturates to a value lower than that of a random pure state for all accessible times. For the dynamics under $\hat{H}_\mathrm{QZE}$, we have set $V=10^4J$ in Eq.~\eqref{eq:effective} for these results.}
	\label{fig:EE} 
\end{figure}

\subsection{Mid-chain entanglement entropy}
We now look at disorder-free localization through the mid-chain entanglement entropy $\mathcal{S}_{L/2}(t)$. Its dynamics is shown in Fig.~\ref{fig:EE}(a,b) for quenches with the faulty Hamiltonian $\hat{H}=\hat{H}_0+\lambda\hat{H}_1+V\hat{H}_W$ using the error in Eq.~\eqref{eq:H1} and for the error-free case when $\lambda=0$. Both cases yield qualitatively identical behavior at large $V$, which is indeed to be expected since the effective Hamiltonian $\hat{H}_\mathrm{QZE}$ is dominated by $V\hat{H}_W$ (at short times mostly) and $\sum_\mathbf{w}\hat{\mathcal{P}}_\mathbf{w}\hat{H}_0\hat{\mathcal{P}}_\mathbf{w}$ (at later times). The dynamics in the presence of errors in Fig.~\ref{fig:EE}(a) shows that the mid-chain entanglement entropy of the unprotected case grows to its maximal possible value, well-approximated by the mid-chain entanglement entropy of a random pure state. Even though in the protected case at finite $V$ the entanglement entropy will always eventually reach this maximal value, this process is delayed linearly in $V$ when the latter is sufficiently large. Specifically, the dynamics is well-reproduced by the effective gauge theory $\hat{H}_\mathrm{QZE}=V\hat{H}_W+\sum_\mathbf{w}\hat{\mathcal{P}}_\mathbf{w}\big(\hat{H}_0+\lambda\hat{H}_1\big)\hat{\mathcal{P}}_\mathbf{w}$ up to a timescale $\propto V/J^2$, just as in the case of the imbalance and superselection-sector projectors. After this timescale, we find that the entanglement entropy grows again before settling into an intermediate plateau, which signifies the emergence of a renormalized gauge theory up to a timescale $\propto V^2/(\lambda^2 J)$, which preserves only the $\mathbb{Z}_2$ gauge symmetry generated by $\hat{G}_j$, but not the enhanced local symmetry of $\hat{H}_\mathrm{QZE}$ associated with $\hat{W}_j$. Indeed, comparing the duration and timescale of this plateau for $V=10^4J$ in Fig.~\ref{fig:EE}(a), we find that they roughly match their counterparts in Figs.~\ref{fig:nonGI_imbalance}(b) for the imbalance and~\ref{fig:projectors}(a) for $\langle\hat{\mathcal{P}}_\mathbf{g}\rangle$.

In the absence of errors, the quench dynamics of the entanglement entropy under $\hat{H}_0$ saturates to a value lower than that of a random pure state, since disorder-free localization is not compromised. Upon adding LPG protection at sufficiently large $V$, the localization at long times is strengthened. Up to times linear in $V$, the dynamics of the entanglement entropy is well-reproduced by the effective gauge theory $\hat{H}_\mathrm{QZE}=V\hat{H}_W+\sum_\mathbf{w}\hat{\mathcal{P}}_\mathbf{w}\hat{H}_0\hat{\mathcal{P}}_\mathbf{w}$. After this timescale, the entanglement entropy grows again and saturates near the value under $\hat{H}_0$.

It is interesting to note in Fig.~\ref{fig:EE} how the entanglement entropy grows faster at very early times in the presence of LPG protection than in the unprotected case. Rigorously, in the large-$V$ limit, the QZE Hamiltonian should involve the protection term itself, in addition to the projector part; see Eq.~\eqref{eq:effective} and Sec.~\ref{sec:QZE}. In the case of the spin-$S$ $\mathrm{U}(1)$ quantum link models \cite{Halimeh2021stabilizingDFL}, the protection term, composed of the local generator $\hat{G}_j$, induces dynamics only in the sectors in which the initial state is prepared, and so it will not lead on its own to any significant increase in the entanglement entropy at early times. In the case of the $\mathbb{Z}_2$ lattice gauge theory, on the other hand, the LPG protection involves strong dynamics at large $V$ in the superselection sectors of the gauge symmetry generated by $\hat{W}_j$. This will lead to a faster growth of entanglement entropy at short times as the involved dynamics is not pure intra-sector dynamics from the viewpoint of the superselection sectors of the $\mathbb{Z}_2$ gauge symmetry generated by $\hat{G}_j$. In a way, this is a small price to pay: adding new local-symmetry sectors due to the LPG protection that the dynamics can now explore leads to a faster entanglement-entropy growth at very early times due to $V\hat{H}_W$, but also brings about greater localization at intermediate to late times, where the projector part $\sum_\mathbf{w}\hat{\mathcal{P}}_\mathbf{w}\hat{H}_0\hat{\mathcal{P}}_\mathbf{w}$ is dominant. Indeed, as we have seen for the imbalance and superselection-sector projectors, the term $V\hat{H}_W$ is inconsequential to their dynamics under $\hat{H}_\mathrm{QZE}$.

\section{Quantum Zeno subspaces}\label{sec:QZE}
To put an analytic footing on the numerical results we have presented in our work, we now employ the quantum Zeno effect (QZE) to build an effective model that can faithfully reproduce the disorder-free localization due to the LPG protection. In the limit of large $V$, the dynamics under the faulty theory $\hat{H}=\hat{H}_0+\lambda\hat{H}_1+V\hat{H}_W$ is limited to the nondegenerate ``decoherence-free'' subspaces of the LPG protection operator $\hat{H}_W$. Let us denote these quantum Zeno subspaces by their projectors $\hat{P}_\alpha$, which will satisfy the relation $\hat{H}_W\hat{P}_\alpha=\epsilon_\alpha\hat{P}_\alpha$, where $\epsilon_\alpha$ are unique for every $\alpha$, i.e., $\epsilon_\alpha=\epsilon_{\alpha}\iff\alpha=\alpha'$. The time-evolution operator $\hat{U}_V(t)=e^{-i\hat{H}t}$ can be shown to be diagonal in the basis of $\hat{H}_W$ in the limit of $V\to\infty$, such that $\big[\hat{U}_{V\to\infty}(t),\hat{P}_\alpha\big]=0$.

To prove this \cite{facchi2002quantum}, let us go into the interaction picture of $\hat{H}_0+\lambda\hat{H}_1$, which we denote by the superscript $\mathrm{I}$. The Schr\"odinger equation then reads
\begin{align}
    i\partial_t\hat{U}_V^\mathrm{I}(t)=V\hat{H}_W^\mathrm{I}(t)\hat{U}_V^\mathrm{I}(t).
\end{align}
In the large-$V$ limit, this is identical to an adiabatic evolution with $V$ corresponding to large time. In the $V\to\infty$ limit, the intertwining property $\hat{U}_{V\to\infty}^\mathrm{I}(t)\hat{P}_\alpha^\mathrm{I}(0)=\hat{P}_\alpha^\mathrm{I}(t)\hat{U}_{V\to\infty}^\mathrm{I}(t)$ is satisfied. This means that $\hat{U}_{V\to\infty}^\mathrm{I}(t)$ maps the Hilbert subspace $\mathcal{H}_{\hat{P}_\alpha^\mathrm{I}(0)}$ of $\hat{P}_\alpha^\mathrm{I}(0)$ exactly into the subspace $\mathcal{H}_{\hat{P}_\alpha^\mathrm{I}(t)}$ of $\hat{P}_\alpha^\mathrm{I}(t)$. In the Schr\"odinger picture, this translates to
\begin{align}
    \ket{\phi(0)}\in\mathcal{H}_{\hat{P}_\alpha}\iff\ket{\phi(t)}\in\mathcal{H}_{\hat{P}_\alpha},
\end{align}
meaning that any state $\ket{\phi(0)}$ initially in the subspace $\mathcal{H}_{\hat{P}_\alpha}$ will undergo dynamics restricted within that subspace when $V\to\infty$. Indeed, the formalism of the adiabatic theorem yields the effective time-evolution operator
\begin{align}
    \hat{U}_{V\to\infty}(t)=e^{-i[V\hat{H}_W+\sum_\alpha\hat{P}_\alpha(\hat{H}_0+\lambda\hat{H}_1)\hat{P}_\alpha]t},
\end{align}
up to an error with an upper bound $\propto tV_0^2L^2/V$ \cite{facchi2004unification}. The energetic term $V_0$ is roughly a linear sum of $J$, $h$, and $\lambda$.

The relation between the quantum Zeno projectors $\hat{P}_\alpha$ and the superselection projectors $\hat{\mathcal{P}}_\mathbf{w}$ and $\hat{\mathcal{P}}_\mathbf{g}$ intimately depends on the choice of the sequence $c_j$ in Eq.~\eqref{eq:LPGprotection}. It may be possible to engineer a sequence such that the spectral decomposition of $\hat{H}_W$ is exactly such that each quantum Zeno subspace with projector $\hat{P}_\alpha$ corresponds to one and only one superselection sector $\hat{\mathcal{P}}_\mathbf{w}$, and vice versa. Equivalently, this means that
\begin{align}\label{eq:compliance}
\hat{H}_W\hat{\mathcal{P}}_\mathbf{w}=\epsilon_\mathbf{w}\hat{\mathcal{P}}_\mathbf{w},\text{ with }\epsilon_\mathbf{w}=\epsilon_{\mathbf{w}'}\iff\mathbf{w}=\mathbf{w}'.
\end{align}
However, in such a case it may be unavoidable to have a highly fine-tuned spatially inhomogeneous sequence $c_j$ that is unfeasible experimentally and scales with $L$. Nevertheless, and as our exact diagonalization results show, a simple translation-invariant sequence such as $c_j=[6(-1)^j+5]/11$ will still offer reliable protection and enhancement of disorder-free localization. Such simple sequences may in general not guarantee the relation~\eqref{eq:compliance}, and then a quantum Zeno subspace will be a union of a few superselection sectors, i.e., $\hat{P}_\alpha=\sum_{\{\mathbf{w};\,\sum_jc_jw_j=\epsilon_\alpha\}}\hat{\mathcal{P}}_\mathbf{w}$. But with the appropriate sequence, all sets $\{\mathbf{w};\,\sum_jc_jw_j=\epsilon_\alpha\}$ will only contain a few sectors $\mathbf{w}$ for each $\alpha$, and, more importantly, each set will have sectors that do not couple up to first order in $\hat{H}_0+\lambda\hat{H}_1$. Once this is satisfied, then in the regime of validity of the quantum Zeno effect, the time-evolution operator can be written directly as
\begin{align}
    \hat{U}_{V\to\infty}=e^{-i\hat{H}_\mathrm{QZE}t},
\end{align}
where the emergent gauge theory, $\hat{H}_\text{QZE}$ given in Eq.~\eqref{eq:effective}, is what we have used in our exact diagonalization results. It is worth noting that $\hat{H}_\text{QZE}$ has an enhanced local symmetry that contains the original $\mathbb{Z}_2$ gauge symmetry generated by $\hat{G}_j$. This enhanced local symmetry is associated with the $\mathbb{Z}_2$ LPG $\hat{W}_j$. This is embodied in the commutation relations $\big[\hat{H}_\text{QZE},\hat{G}_j\big]=\big[\hat{H}_\text{QZE},\hat{W}_j\big]=0,\,\forall j$.

So far, we have understood how the gauge protection $V\hat{H}_W$ leads to quantum Zeno subspaces that are roughly equivalent to the superselection sectors $\mathbf{w}$, thereby suppressing transitions between these sectors up to timescales $\propto V/(V_0L)^2$. But in general, the sectors $\mathbf{w}$ and $\mathbf{g}$ are not the same since $\hat{W}_j\neq\hat{G}_j$. However, the \textit{occupied} sectors $\mathbf{w}$ and $\mathbf{g}$ that constitute the initial superposition state $\ket{\psi^z_0}$ are equivalent (see Tables~\ref{Table_G_psi} and~\ref{Table_W_psi}), and at large $V$ the sectors $\mathbf{w}$ are actual local-symmetry sectors since the dynamics is effectively propagated by $\hat{H}_\text{QZE}$. Despite being equivalent, the superselection sectors $\mathbf{w}$ and $\mathbf{g}$ generally have different background charges, which leads to a greater effective disorder in the ensuing dynamics thus enhancing disorder-free localization. Interestingly, the local symmetry due to $\hat{W}_j$ is subjected to the errors $\hat{H}_0+\lambda\hat{H}_1$, while the gauge symmetry due to $\hat{G}_j$ is only subjected to the errors $\lambda\hat{H}_1$. This is the reason why the timescale over which the local symmetry associated with $\hat{W}_j$ is preserved is shorter than that for the $\mathbb{Z}_2$ gauge symmetry. Let us consider the $\mathbb{Z}_2$ sector $\mathbf{g}=(+1,+1,-1,-1)$. This corresponds to the $\hat{W}_j$ sectors $\mathbf{w}=(+1,+1,-1,-1)$, $\mathbf{w}'=(+1,+1,-1,+3)$, $\mathbf{w}''=(+1,+1,+3,-1)$, and $\mathbf{w}'''=(+1,+1,+3,+3)$. However, when the system is prepared in $\ket{\psi^z_0}$, only the sector $\mathbf{w}=(+1,+1,-1,-1)$ is occupied at $t=0$. The $\mathbb{Z}_2$ Hamiltonian $\hat{H}_0$ can only lead to transitions between $\mathbf{w}$, $\mathbf{w}'$, $\mathbf{w}''$, and $\mathbf{w}'''$, but no other $\hat{W}_j$ sectors, as then this would mean transitions between $\mathbf{g}$ and other $\mathbb{Z}_2$ sectors, which is not possible since $\big[\hat{H}_0,\hat{G}_j\big]=0,\,\forall j$. As such, we see that after the timescale $\propto V/J^2$ the local symmetry due to $\hat{W}_j$ is broken by $\hat{H}_0$ in a \textit{special} way that still guarantees the conservation of the $\mathbb{Z}_2$ gauge symmetry up to the timescale $\propto V^2/(\lambda^2 J)$. However, at long enough times, $\lambda\hat{H}_1$ will be the dominant error, and then both local symmetries will be completely broken in the dynamics.

\begin{figure}[t!]
	\centering
	\includegraphics[width=.48\textwidth]{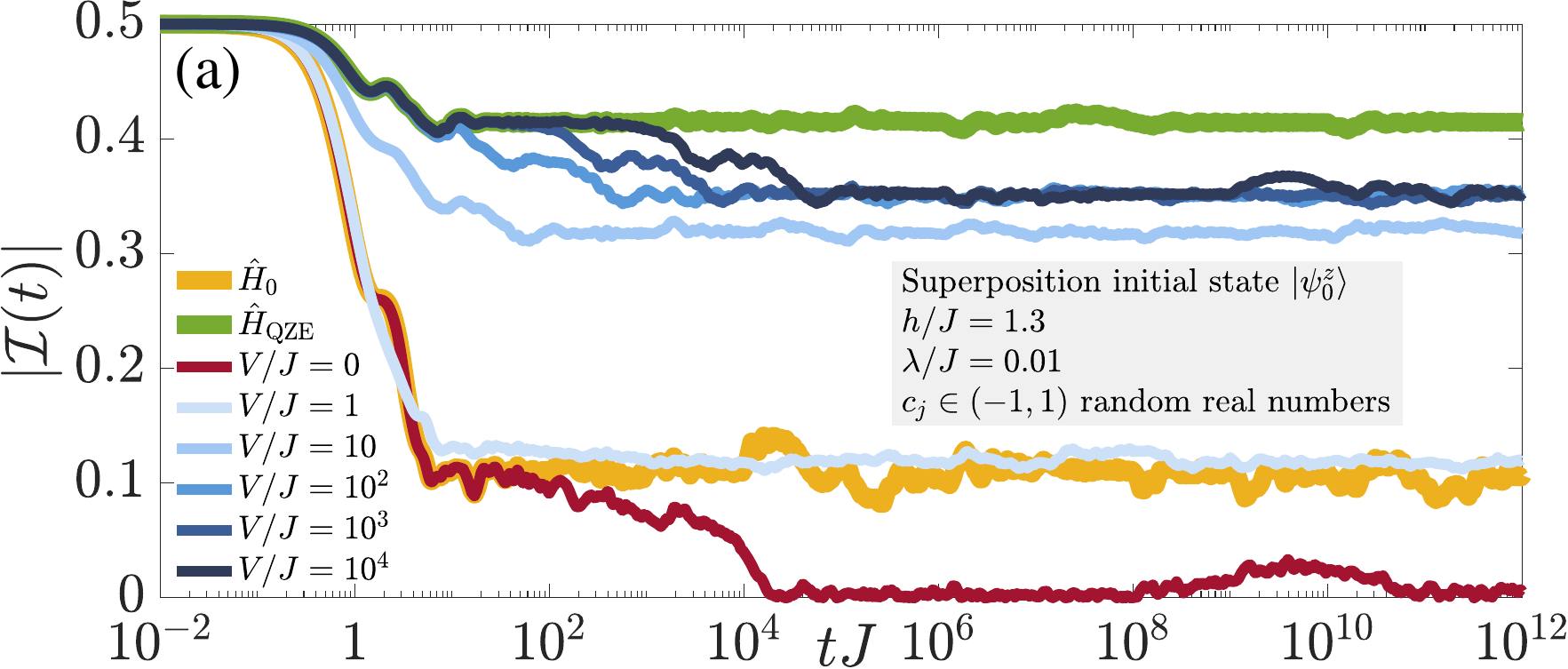}\\
	\vspace{0.1cm}
	\includegraphics[width=.48\textwidth]{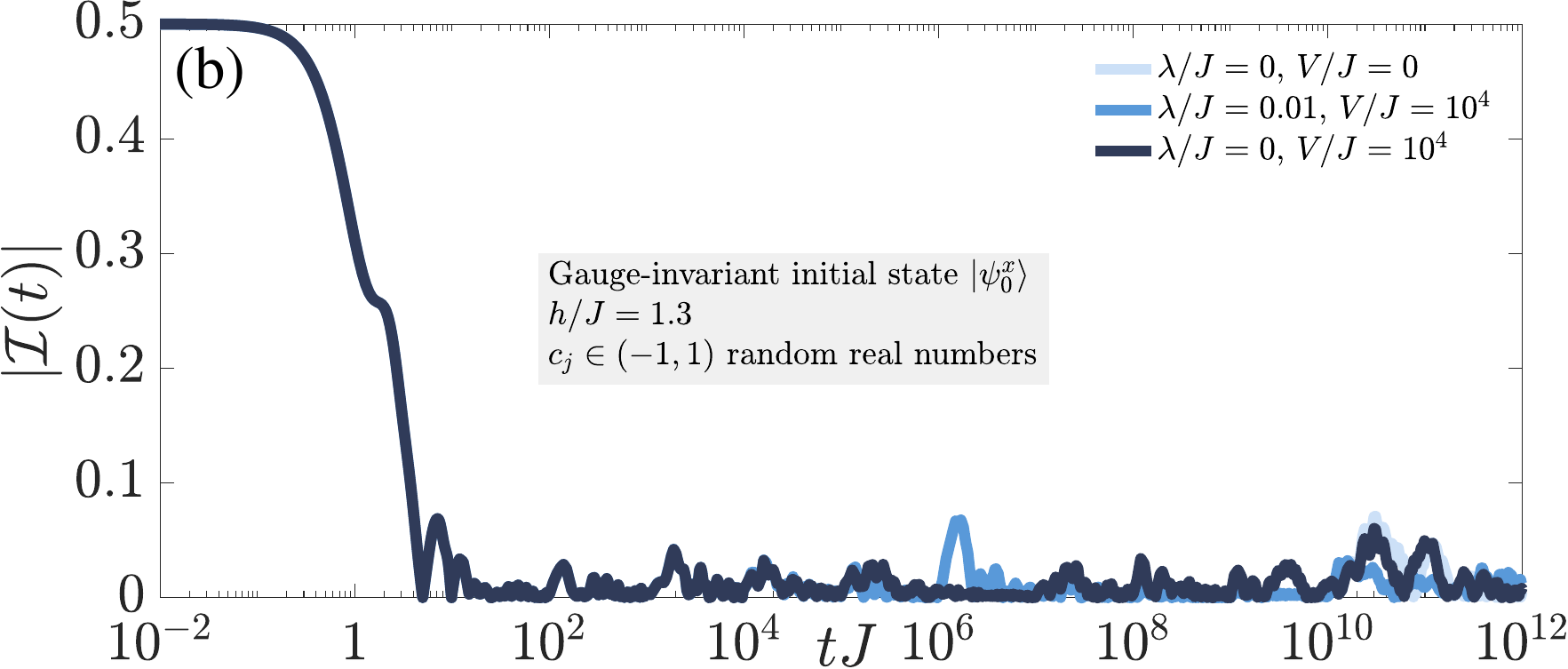}
	\caption{(Color online). LPG protection with a translation-noninvariant sequence $c_j\in(-1,1)$ of random real numbers. (a) When starting in the superposition initial state $\ket{\psi^z_0}$, the LPG protection at sufficiently large $V$ leads to localized dynamics that is well-reproduced by the same effective gauge theory $\hat{H}_\mathrm{QZE}=\sum_\mathbf{w}\hat{\mathcal{P}}_\mathbf{w}\big(\hat{H}_0+\lambda\hat{H}_1\big)\hat{\mathcal{P}}_\mathbf{w}$ that emerges in the case of a translation-invariant $c_j$, up to a timescale polynomial in $V$. However, the renormalized gauge theory that emerges after this timescale, which conserves only the $\mathbb{Z}_2$ gauge symmetry, persists over all accessible evolution times in exact diagonalization, which is not the case when $c_j$ is translation-invariant, where the imbalance eventually thermalizes to zero, see Fig.~\ref{fig:nonGI_imbalance}(b). (b) Starting in the gauge-invariant initial state $\ket{\psi^x_0}$, LPG protection with a random $c_j$ does not protect localized behavior regardless of $\lambda$ and $V$.}
	\label{fig:random} 
\end{figure}

\begin{table}[t!]
	\centering
	\begin{tabular}{|| c || c | c ||}
		\hline
		 $\mathbf{g}=(g_1,g_2,g_3,g_4)$ & $\big\lvert\big\lvert\hat{P}_\mathbf{g}\hat{H}_0\hat{P}_\mathbf{g}\big\rvert\big\rvert$ & $\big\lvert\big\lvert\sum_\mathbf{w}\hat{P}_\mathbf{g}\hat{P}_\mathbf{w}\hat{H}_0\hat{P}_\mathbf{w}\hat{P}_\mathbf{g}\big\rvert\big\rvert$ \\ [0.5ex] 
		\hline\hline
		$(-1,-1,-1,-1)$ & $2.8914$ & $0.6$\\ 
		\hline
		$(-1,-1,+1,+1)$ & $3.1253$ & $1.344$\\ 
		\hline
		$(-1,+1,-1,+1)$ & $3.284$ & $1.2$\\ 
		\hline
		$(-1,+1,+1,-1)$ & $3.1253$ & $1.344$\\ 
		\hline
		$(+1,-1,-1,+1)$ & $3.1253$ & $1.344$\\ 
		\hline
		$(+1,-1,+1,-1)$ & $3.284$ & $1.2$\\ 
		\hline
		$(+1,+1,-1,-1)$ & $3.1253$ & $1.344$\\ 
		\hline
		$(+1,+1,+1,+1)$ & $2.8914$ & $2.8914$\\  [1ex] 
		\hline
	\end{tabular}
	\caption{$2$-norm of the projections of $\hat{H}_0$ and $\sum_\mathbf{w}\hat{\mathcal{P}}_\mathbf{w}\hat{H}_0\hat{\mathcal{P}}_\mathbf{w}$ onto the superselection sectors $\mathbf{g}$ constituting the superposition initial state $\ket{\psi^z_0}$ (see Table~\ref{Table_G_psi}). The nonzero renormalized norms show that LPG protection does not trivially freeze dynamics, but that rather in its presence there is renormalized dynamics in the relevant sectors $\mathbf{g}$.}
	\label{TableG}
\end{table}

\section{Discussion}\label{sec:discussion}
It is worth discussing further details regarding the nature of the LPG protection sequence and the local symmetry arising due to the LPG protection, which we do in the following.

\subsection{LPG protection sequence}
As explained in Sec.~\ref{sec:quench} and derived in Sec.~\ref{sec:QZE}, an effective gauge theory emerges dynamically based on the principle of quantum Zeno subspaces \cite{facchi2002quantum}. This requires an appropriate choice of the sequence $c_j$ that penalizes the majority of transitions between different superselection sectors. In principle, if $c_j$ is completely random, then this will give rise to optimal performance in LPG protection, and in fact the system will not thermalize for all accessible evolution times even in the presence of gauge-breaking errors when starting in a superposition initial state such as $\ket{\psi^z_0}$, as shown in Fig.~\ref{fig:random}(a). Even though the dynamics is qualitatively identical to the case of the translation-invariant $c_j$ employed in Fig.~\ref{fig:nonGI_imbalance}(b) up to a timescale $\propto V/J^2$, where the effective gauge theory $\hat{H}_\mathrm{QZE}=\sum_\mathbf{w}\hat{\mathcal{P}}_\mathbf{w}\big(\hat{H}_0+\lambda\hat{H}_1\big)\hat{\mathcal{P}}_\mathbf{w}$ emerges, after this timescale the dynamics is fundamentally different, as the second plateau, which only lasts up to a timescale $\propto V^2/(\lambda^2 J)$ for $c_j=[6(-1)^j+5]/11$, persists indefinitely for a random sequence $c_j$ at a value not predicted by the corresponding thermal ensemble. Even though one may no longer legitimately be able to claim \textit{disorder-free} localization with a random $c_j$, this still would not be typical disorder-MBL, since a random $c_j$ does not allow LPG protection to create disorder when starting in $\ket{\psi^x_0}$, for example, as shown in Fig.~\ref{fig:random}(b).

Furthermore, we emphasize that LPG protection does not trivially ``freeze'' dynamics in the superselection sectors $\mathbf{g}$. All the sectors constituting the superposition of $\ket{\psi^z_0}$ will still undergo dynamics, as illustrated in Table~\ref{TableG}. The projectors $\hat{\mathcal{P}}_\mathbf{g}$ of the relevant superselection sectors $\mathbf{g}$ show a finite $2$-norm when acting on $\hat{H}_0$. We see that in the large-$V$ limit where the protected dynamics is effectively propagated by $\sum_\mathbf{w}\hat{\mathcal{P}}_\mathbf{w}\hat{H}_0\hat{\mathcal{P}}_\mathbf{w}$, the $2$-norm of the projected parts of this term by $\hat{\mathcal{P}}_\mathbf{g}$ are nonzero but just renormalized.

\subsection{Enriched local symmetry due to the LPG}
Even though the concept of LPG protection has already been introduced in Ref.~\cite{Halimeh2021stabilizing}, it is important to stress here that the enriched local symmetry due to LPG protection introduced in this work could not have been present in the setup of Ref.~\cite{Halimeh2021stabilizing}. The reason is that in Ref.~\cite{Halimeh2021stabilizing} the main objective is to stabilize dynamics within a single target gauge superselection sector, for which the LPG is engineered such that it acts identically to the full local generator within that target sector. In other words, the LPG and the full local generator are impossible to distinguish in the target sector, and therefore one cannot resolve any enriched local symmetry in that setup; see the case in Fig.~\ref{fig:schematic}(b) where the yellow block (sector due to the full local generator) and the green block (sector due to the LPG) fully overlap. Indeed, the effective Zeno Hamiltonian when working only within a target superselection sector is \cite{Halimeh2021stabilizing}
\begin{align}\label{eq:effective_reduced}
    \hat{H}_\text{QZE}^\text{tar}=\hat{H}_0+\lambda\hat{\mathcal{P}}_{\mathbf{g}^\text{tar}}\hat{H}_1\hat{\mathcal{P}}_{\mathbf{g}^\text{tar}},
\end{align}
which is a reduced version of Eq.~\eqref{eq:effective}, and where $\mathbf{g}^\text{tar}=\mathbf{w}^\text{tar}$ is the target sector. Starting in an initial state $\ket{\psi^\text{tar}_0}$ in the target sector, one cannot distinguish as to whether its dynamics is propagated by Eq.~\eqref{eq:effective} or Eq.~\eqref{eq:effective_reduced}, simply because $\hat{\mathcal{P}}_\mathbf{g}\ket{\psi^\text{tar}_0}\neq0\iff\mathbf{g}=\mathbf{g}^\text{tar}$. On the other hand, starting in a superposition initial state as we have done in the main results of this work, will lead to fundamentally different dynamics whether the quench Hamiltonian is Eq.~\eqref{eq:effective} or Eq.~\eqref{eq:effective_reduced}, whereby only in the former case can enhanced disorder-free localization emerge due to the enriched local symmetry.

This highlights the intimate connection between gauge invariance and disorder-free localization. Restricting to one superselection sector will not resolve the local symmetry of the model, as no information is available on the structure of the other gauge-symmetry sectors. But a superposition initial state allows resolving the full structure of the local symmetry, and this will therefore lead to localized dynamics, where an emergent disorder over the background charges associated with the superselection sectors arises, even though the system and the initial state themselves are translation-invariant and disorder-free. In the case of LPG protection, we have shown how this local symmetry is enhanced with respect to that of the original model. In this case, the initial state is a superposition over a larger number of superselection sectors, thereby effecting a greater emergent disorder over the associated background charges, and hence leading to stronger disorder-free localization.

\section{Experimental proposal}\label{sec:exp}
\begin{figure*}[t]
\includegraphics[width=0.95\textwidth]{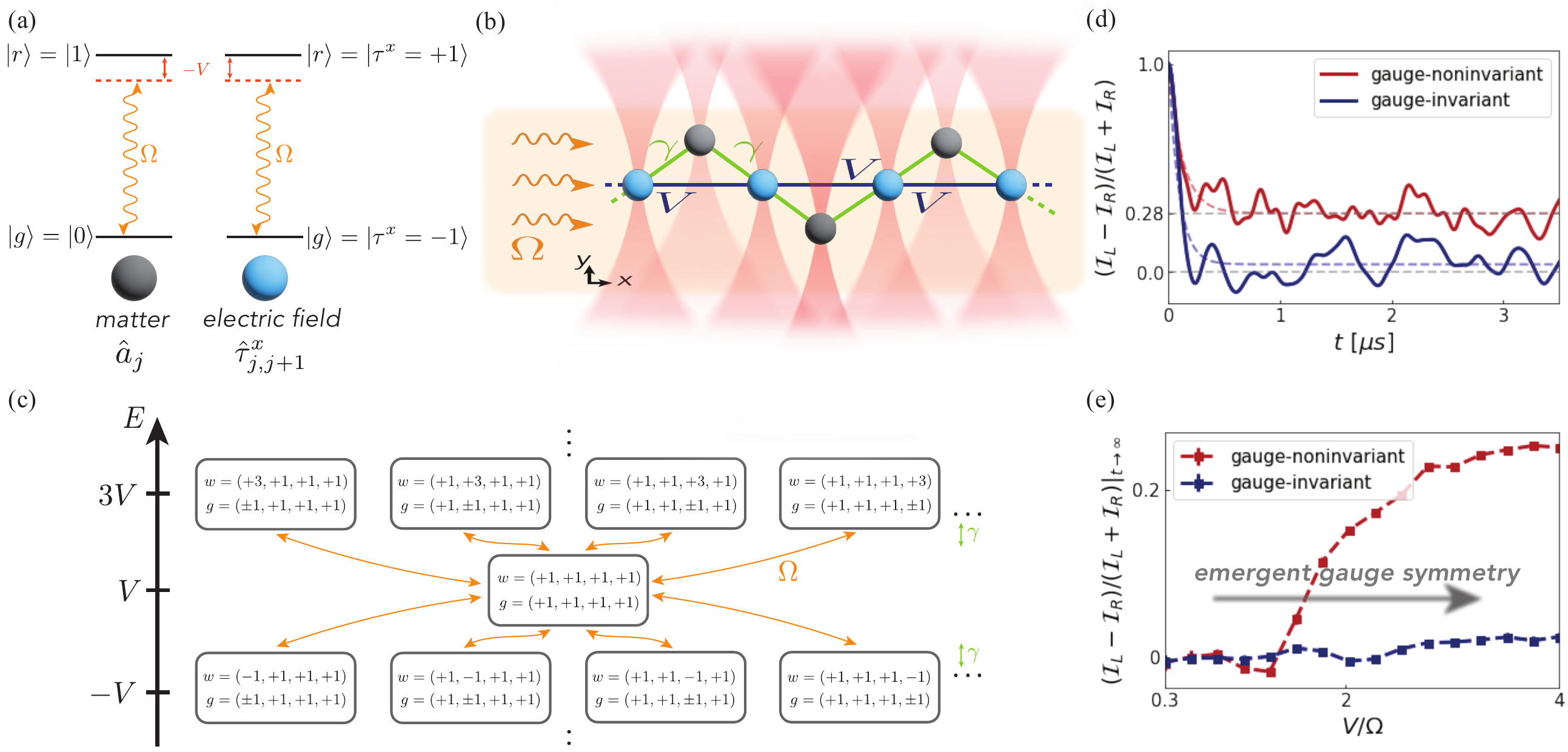}
\caption{(Color online). (a) The matter and gauge degrees of freedom can be mapped onto qubits, where the two energy levels are, e.g., given by the ground and Rydberg states in an ultracold-atom setup. Likewise, superconducting qubits can also be used in our experimental proposal. Here, gray (red) circles depict the matter (electric) fields on the sites (links) of the one-dimensional lattice. The qubits have to be driven at strength $\Omega$ and detuning~$-V$. We note that the detuning is chosen to have the same strength as the LPG protection. (b) In the case of atomic tweezer arrays, the density-density interaction can be controlled by the inter-atomic distance~$r$ and scales as~$r^{-6}$. The shown setup is a possible configuration to implement Hamiltonian~\eqref{eq:micH}. In our proposal, we require~$V = 2/3\gamma$. (c) The LPG protection terms~$\propto V$ separate the different $\hat{W}_j$ sectors energetically. Further, the terms~$\propto \gamma$ stabilize the system against errors, where two sectors are lifted up and down in energy. The drive~$\Omega$ couples between different sectors and hence induces the desired dynamics within the energy sectors in higher-order perturbation theory. (d) We plot the raw (temporally nonaveraged) imbalance for different initial states (solid lines) and fit the result with an exponential function (dashed lines). The gauge-noninvariant superposition state $\ket{\psi^z_0}$ shows a prethermal localizing behavior which is distinctly different from the thermalization of the gauge-invariant initial state to zero imbalance. We use experimentally realistic parameters $\gamma = 2\pi \times 4.5\,\mathrm{MHz}$, $V=2\pi \times 3\,\mathrm{MHz}$ and $\Omega = 2\pi \times 1\,\mathrm{MHz}$, which leads to interesting dynamics within experimentally feasible timescales of~$\approx 3 \mu s$. (e) As in (d), we fit the imbalance for different driving strengths~$V/\Omega$ and extract the $t\rightarrow \infty$ value. We find two different regimes with and without disorder-free localization that emphasizes the emergent gauge symmetry in the limit of weak driving~$V/\Omega \gtrsim 1$.}
\label{fig:RydbergScheme}
\end{figure*}

In this Section, we propose a readily feasible experimental scheme to observe disorder-free localization in an analog quantum simulation platform. In the following, we will focus on Rydberg atom arrays~\cite{Browaeys_review}, although the proposed scheme can similarly be adapted for superconducting qubits~\cite{Kjaergaard2020}.
The difficulty lies in the implementation of a gauge-invariant Hamiltonian such as Eq.~\eqref{eq:H0}. To circumvent the explicit construction of involved $\mathbb{Z}_2$ gauge-invariant dynamical schemes \cite{Schweizer2019, homeier2020mathbbz2}, which involve three-body interactions, we only implement two-body LPG terms~\eqref{eq:LPG} in the Hamiltonian while driving the system, as we will elaborate in the following.
The key idea is that strong LPG terms create well-defined energy subspaces, which are the symmetry sectors, and integrating out the weak drive yields the following two interactions:
1) States within the same symmetry sector are energetically on resonance but not coupled directly by single-body driving terms. Nevertheless, virtually coupling to other sectors can induce effective interactions such as Eq.~\eqref{eq:H0} after a Schrieffer-Wolff transformation.
2) Two states in different energy sectors can be coupled by single-body driving terms but are not on resonance. Therefore, these effective couplings, which can be considered as gauge symmetry-breaking terms, are suppressed by strong LPG terms.

Here, the goal is not to implement the exact Hamiltonian~\eqref{eq:H0} but instead we only argue that the dynamics is governed by an effective Hamiltonian~$\hat{H}_\mathrm{eff}$ with~$[\hat{H}_\mathrm{eff},\hat{W}_j]=0,~\forall j$. Regardless of the microscopic details of~$\hat{H}_\mathrm{eff}$, we find qualitatively the same disorder-free localizing behavior as discussed above.

For the experimental setup, both matter and link degrees of freedom are individually mapped onto qubits realized by single atoms or superconducting qubits. The matter site has two local basis states described by the unoccupied state $\ket{g}_j$ and the occupied state $\ket{r}_j = \hat{a}^\dagger_j \ket{g}_j$, where $\ket{g}_j$ ($\ket{r}_j$) is the ground (Rydberg) state at site $j$, see Fig.~\ref{fig:RydbergScheme}(a).
Similarly, the local Hilbert space on the link between matter sites $j$ and $j+1$ is described by the two electric-field configurations $\ket{g}_{j,j+1}$ for~$\tau^x_{j,j+1}=-1$ and $\ket{r}_{j,j+1} = \hat{a}^\dagger_{j,j+1} \ket{g}_{j,j+1}$ for~$\tau^x_{j,j+1}=+1$, and therefore the electric field configurations can be written as hard-core bosonic occupations of the link qubit $\hat{n}_{j,j+1}=\hat{a}^\dagger_{j,j+1}\hat{a}_{j,j+1} = \big(\hat{\tau}^x_{j,j+1} + \hat{\mathds{1}}_{j,j+1}\big)/2$.

The microscopic Hamiltonian we propose to implement is given by
\begin{align}\nonumber
    \hat{H}_{\mathrm{mic}} = \sum_j \bigg[&V \hat{n}_{j-1,j}\hat{n}_{j,j+1} 
    + \gamma\hat{n}_j \big( \hat{n}_{j-1,j} + \hat{n}_{j,j+1} \big)\\\nonumber
    &+\big(h-V\big) \hat{n}_{j,j+1} +\bigg(\mu+\frac{V}{2}\bigg)\hat{n}_j\\\label{eq:micH}
    &+ \Omega\big(\hat{a}_j + \hat{a}_{j,j+1} + \mathrm{H.c.} \big)\bigg].
\end{align}
The terms with coupling strengths~$\propto V$ comprise the LPG protection terms in the qubit basis. The density-density interaction terms can be implemented by density dependent Rydberg--Rydberg interactions and controlled by their relative distance, e.g., in tweezer atom arrays; see Fig.~\ref{fig:RydbergScheme}(b).
Additionally, we require an on-site drive~$\Omega$ coupling the ground and Rydberg states, $\ket{g} \leftrightarrow \ket{r}$, for each qubit.
In the rotating frame of the qubits, the drive yields to the creation and annihilation of qubit excitations while the detuning from the bare qubit frequency gives rise to single-body terms~$\propto \hat{n}_j$ and~$\propto \hat{n}_{j,j+1}$.
Moreover, $\mu$ is the chemical potential for matter excitations and $h$ is the electric field.

Note that the LPG term in Eq.~\eqref{eq:LPG} has the form (for $c_j = g_j=+1,~\forall j$)
\begin{align}\nonumber
    &V\sum_{j} \left( \hat{\tau}^x_{j-1,j}\hat{\tau}^x_{j,j+1} +2\hat{n}_j  \right)\\
    = &2V\sum_{j}\left(2
    \hat{n}_{j-1,j}\hat{n}_{j,j+1} -2\hat{n}_{j,j+1} +\hat{n}_j \right) +\mathrm{const.} \label{eq:LPGinRy}
\end{align}
Comparing Eqs.~\eqref{eq:micH} and~\eqref{eq:LPGinRy} shows that the interaction strength has been adapted ($4V \rightarrow V$) in order to set the Rydberg--Rydberg interaction as the natural energy scale in the system.

Additionally, we introduce interactions between electric-field and matter qubits with strength~$\gamma$.
On the one hand, this density-density interaction~$\gamma$ ensures that the desired inter-site correlation is built up because for~$\gamma=0$ the matter and gauge fields would be decoupled.
On the other hand, we want to work with~$c_j=+1,~\forall j$, which leads to more fragility of the~$w_j=g_j=+1$ sector against errors.
This can be seen by considering the eigenvalues~$w_j\in \{ -1,+1,+3\}$ of the LPG term.
Assume the system is initially prepared in a state with~$w_i=w_j=+1$ for two sites~$i$ and~$j$.
Then a gauge-symmetry breaking error can resonantly couple to a sector with~$w_i=-1$ and~$w_j=+3$, which have the same energy under LPG protection.
This resonance can be lifted by choosing~$\gamma \neq 0$ and large~$\gamma \gg \Omega$.
In general, this would also lead to a splitting of the~$w_j=+1$ manifold.
However, by choosing~$\mu=-\gamma$ we can compensate the undesired splitting of the~$w_j=+1$ manifold while the resonance with other sectors caused by gauge-symmetry breaking errors is effectively suppressed because only the~$w_j=+3$ is modified by~$\gamma$ terms as indicated in Fig.~\ref{fig:RydbergScheme}(c).

Moreover, for experimental simplicity it is convenient to have uniform driving strength and detuning.
Hence, the condition $(h-V)=(\mu+V/2)$ shall be fulfilled, which requires $V/\gamma=2/3$ for~$h=0$ and~$\mu=-\gamma$.

The main advantage of our proposed scheme is that no direct implementation of gauge-invariant three-body coupling terms~$\propto \big(\hat{a}^\dagger_j \hat{\tau}^z_{j,j+1} \hat{a}_{j+1} +\mathrm{H.c.}\big)$ is required. To relate Hamiltonian~\eqref{eq:micH} to our above findings, we consider the following: the LPG protection terms enforce a large energy splitting between the local-symmetry superselection sectors associated with~$\hat{W}_j$, see Fig.~\ref{fig:RydbergScheme}(c).
In the limit of a weak local on-site drive~$\Omega \ll V,\gamma$ the gauge-invariant coupling terms in the Hamiltonian are then induced in third-order perturbation theory. Therefore, the perturbative dynamics to leading order in~$\Omega/V$, which we will denote by the Hamiltonian $\hat{H}_\mathrm{eff}$, is enforced to act within the symmetry sectors given by~$\hat{W}_j$, i.e., $\big[\hat{H}_\mathrm{eff}, \hat{W}_j \big] =0,\,\forall j$. From relation~\eqref{eq:relation}, this automatically means that $\big[\hat{H}_\mathrm{eff}, \hat{G}_j \big] =0,\,\forall j$. 

As shown in Fig.~\ref{fig:RydbergScheme}(c), the drive~$\Omega$ couples between $\hat{W}_j$ sectors that are separated in energy by~$\mathcal{O}(V,\gamma)$.
Hence, dispersive energy shifts~$\propto \hat{n}_j$ and~$\propto \hat{\tau}^x_{j,j+1}$ arise in second-order perturbation theory with strengths~$\mathcal{O}(\Omega^2 V^{-1},\Omega^2\gamma^{-1})$.
Moreover, the desired three-body coupling terms only appear in third-order processes and thus their coupling strength is of order~$\mathcal{O}(\Omega^3 V^{-2},\Omega^3\gamma^{-2},\Omega^3V^{-1}\gamma^{-1})$. Additionally, error terms of the same order as the gauge-invariant three-body couplings are induced leading to ergodic behavior for long times since the emergent local symmetry is violated.
Because the gauge-symmetry breaking terms are nonresonant processes, we expect the amplitudes of the error terms to be much smaller than the gauge-invariant effective couplings.
Moreover, as we have discovered in Secs.~\ref{sec:quench} and~\ref{sec:QZE}, the presence of strong LPG protection terms not only stabilizes but also enhances disorder-free localization. This enhancement makes it possible to observe this phenomenon on experimentally relevant timescales.

In the following, we want to consider a regime of equal drive and equal detuning on all matter and link sites, i.e., in particular $V/\gamma$, $\mu/\gamma = -1$ and~$h=0$.
We emphasize that this regime is appealing for experimental purposes since, e.g., in Rydberg atom arrays only a single driving laser with fixed detuning and strength for all qubits is required.
Figure~\ref{fig:RydbergScheme}(b) shows a possible arrangement of Rydberg atoms in real-space.
Here, we neglect long-range interactions beyond the ones in Hamiltonian~\eqref{eq:micH} since the interactions decay as~$r^{-6}$ between two Rydberg atoms.

We study the quench dynamics for a system of $L=4$ matter sites and $L=4$ gauge links with periodic boundary conditions using exact diagonalization.
The system is initialized in the gauge-(non)invariant product state~$\ket{\psi_0^x}$ ($\ket{\psi_0^z}$), shown in Fig.~\ref{fig:InitialStates}, and then time-evolved under Hamiltonian~\eqref{eq:micH} in the limit of weak driving~$\Omega/V= 1/3$.
In an experimental setup, the initial states can be prepared by rotating individual qubits into the product states $\ket{r}$ or $(\ket{g}+\ket{r})/\sqrt{2}$ at time~$t=0$. Explicitly, the initial states take the form
\begin{subequations}
\begin{align}
    \ket{\psi_0^x}&=\ket{g}_{1,2}\bigotimes_{j=2,3,4}\ket{r}_{j,j+1}\bigotimes_{j=1,2}\ket{r}_j\ket{g}_{j+2},\\
    \ket{\psi_0^z}&=\bigotimes_{j=1,\ldots,4}\frac{\ket{g}_{j,j+1}+\ket{r}_{j,j+1}}{\sqrt{2}}\bigotimes_{j=1,2}\ket{r}_j\ket{g}_{j+2}.
\end{align}
\end{subequations}

In Fig.~\ref{fig:RydbergScheme}(d), the temporally nonaveraged imbalance~\cite{Choi2016} $(\mathcal{I}_L - \mathcal{I}_R)/(\mathcal{I}_L + \mathcal{I}_R)$ is plotted using experimentally realistic parameters~\cite{Bernien2017}.
Here, $\mathcal{I}_L$ ($\mathcal{I}_R$) measures the occupation of matter sites on the left, $j=1,2$, (right, $j=3,4$) half of the system (see Fig.~\ref{fig:InitialStates}).
The proposed scheme clearly shows a localization of the domain wall for the gauge-noninvariant superposition state $\ket{\psi^z_0}$ while the gauge-invariant initial state $\ket{\psi^x_0}$ quickly delocalizes the domain wall across the entire system, leading to a vanishing imbalance in congruence with thermalization.

To emphasize the importance of a gauge symmetry for disorder-free localization and how that gauge symmetry arises in our perturbative scheme, we compare the system in the weak and strong driving regimes, i.e., in a regime with and without emergent local symmetry, respectively.
To this end, we fit the imbalance and plot its extracted prethermal steady state value for different driving strengths~$V/\Omega$, as shown in Fig.~\ref{fig:RydbergScheme}(e).
In the limit of a strong drive,~$V/\Omega \lesssim 1$, the system has no gauge symmetry and we find that both initial states thermalize with a vanishing imbalance. When the driving strength is decreased,~$V/\Omega \gtrsim 1$, an emergent local symmetry from LPG protection governs the dynamics of the system. The dynamics then distinguishes between the two initial states, and the gauge-noninvariant superposition state $\ket{\psi^z_0}$ leads to a localization of the domain wall on experimentally relevant timescales, while $\ket{\psi^x_0}$ will still result in a vanishing imbalance. The emergent gauge structure with nontrivial dynamics in the weak driving regime is consistent with our picture that well-defined energy subspaces are required to suppress error terms while gauge-invariant dynamics $\hat{H}_\mathrm{eff}$ is induced by virtual processes.

\section{Conclusions and outlook}\label{sec:conc}
In this work, we have extended the concept of gauge protection based on local pseudogenerators to the phenomenon of disorder-free localization in $\mathbb{Z}_2$ lattice gauge theories. This type of protection involves a translation-invariant alternating sum of the local pseudogenerators, which suppresses transitions between different superselection sectors based on the quantum Zeno effect up to timescales polynomial in the protection strength. This preserves localized behavior over these timescales and even enhances it due to the dynamical emergence of an enhanced local symmetry associated with the local pseudogenerator, and which contains the $\mathbb{Z}_2$ gauge symmetry of the ideal theory. The initial state that is a superposition over the superselection sectors of the original $\mathbb{Z}_2$ gauge symmetry is also a superposition over the superselection sectors of the emergent local symmetry. Due to the local-pseudogenerator protection scheme, this leads to a greater effective disorder over superselection sectors, thereby creating stronger localization in the dynamics.

We have provided numerical results from exact diagonalization showing clear protection of disorder-free localization based on local-pseudogenerator protection through the quench dynamics of the imbalance, superselection-sector projectors for both the original $\mathbb{Z}_2$ gauge symmetry and the emergent local symmetry, and the mid-chain entanglement entropy. All these results indicate clear timescales over which disorder-free localization is stabilized and enhanced.

Given the experimental feasibility of the local pseudogenerator, this makes the prospect of realizing an experiment exhibiting stable disorder-free localization a realistic one. We have therefore provided a proposal for such an experiment using Rydberg atoms, where the local-pseudogenerator protection terms are naturally implemented through Rydberg interactions. Driving between the ground and Rydberg states, gauge-invariant dynamics is then perturbatively induced. When starting in an initial state that is a superposition of superselection sectors, experimentally feasible parameters will then give rise to disorder-free localization in the dynamics of the imbalance within experimentally accessible lifetimes. Vice versa, it appears conceivable to introduce local pseudogenerators to further protect topologically ordered systems that can be realized experimentally and already feature emergent discrete gauge theories. This could lead to an enhanced robustness of the corresponding quantum memories.

Even though in this work we have applied our protection scheme to $\mathbb{Z}_2$ lattice gauge theories, we emphasize that our method can be generalized to other Abelian gauge theories. Indeed, a local pseudogenerator can be engineered for any gauge-symmetry generator such that it acts identically to the latter within only a chosen target sector. This construction is not restricted to local generators of $\mathbb{Z}_2$ gauge symmetries, and hence our results can be readily extended.

Several immediate future directions emerge from this work. It would be interesting to see how well linear gauge protection in general will work in higher dimensions. Disorder-free localization has been shown to exist also in $(2+1)$-dimensions \cite{karpov2021disorder}, and our scheme may be a viable way to stabilize it.

With regards to dynamically emergent symmetries due to local-pseudogenerator protection, it would be interesting to investigate whether the concepts we have introduced in this work can be extended to non-Abelian gauge theories, which are currently of great interest to implement in systems of quantum synthetic matter \cite{aidelsburger2021cold}.
              
\begin{acknowledgments}
J.C.H.~is grateful to Ian P.~McCulloch for a thorough reading of and valuable comments on our manuscript. The authors are grateful to Monika Aidelsburger, Debasish Banerjee, Sepehr Ebadi, Arkady Fedorov, Ognjen Markovi\'{c}, Ian P.~McCulloch, and Christian Schweizer for fruitful discussions. This work is part of and supported by Provincia Autonoma di Trento, the ERC Starting Grant StrEnQTh (project ID 804305), the Google Research Scholar Award ProGauge, and Q@TN — Quantum Science and Technology in Trento.  H.Z.~acknowledges support from a Doctoral-Program Fellowship of the German Academic Exchange Service (DAAD). This research was funded by the Deutsche Forschungsgemeinschaft (DFG, German Research Foundation) under Germany's Excellence Strategy -- EXC-2111 -- 390814868 and via DFG Research Unit FOR 2414 under project number 277974659. This project has received funding from the European Research Council (ERC) under the European Union’s Horizon 2020 research and innovation programm (Grant Agreement no 948141). We acknowledge support from the Imperial--TUM flagship partnership. L.H. acknowledges support from the Studienstiftung des deutschen Volkes. A.B.~acknowledges support from the National Science Foundation (NSF) through a grant for the Institute for Theoretical Atomic, Molecular, and Optical Physics at Harvard University and the Smithsonian Astrophysical Observatory.
\end{acknowledgments}

\appendix
\section{Supporting numerical results}\label{app:supp}
Here, we provide supplemental numerical evidence that support the main conclusions of our work.

\subsection{Different error terms}
\begin{figure}[t!]
	\centering
	\includegraphics[width=.48\textwidth]{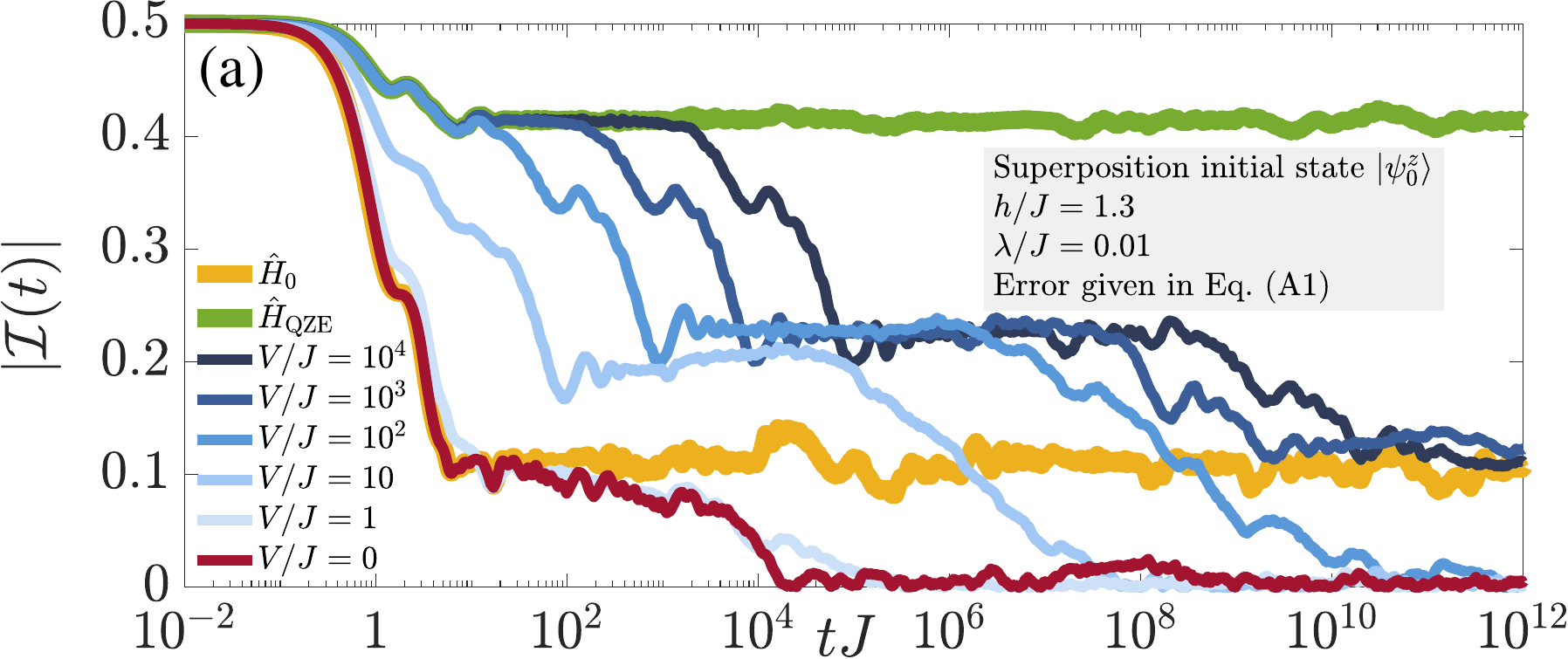}\\
	\vspace{0.1cm}
	\includegraphics[width=.48\textwidth]{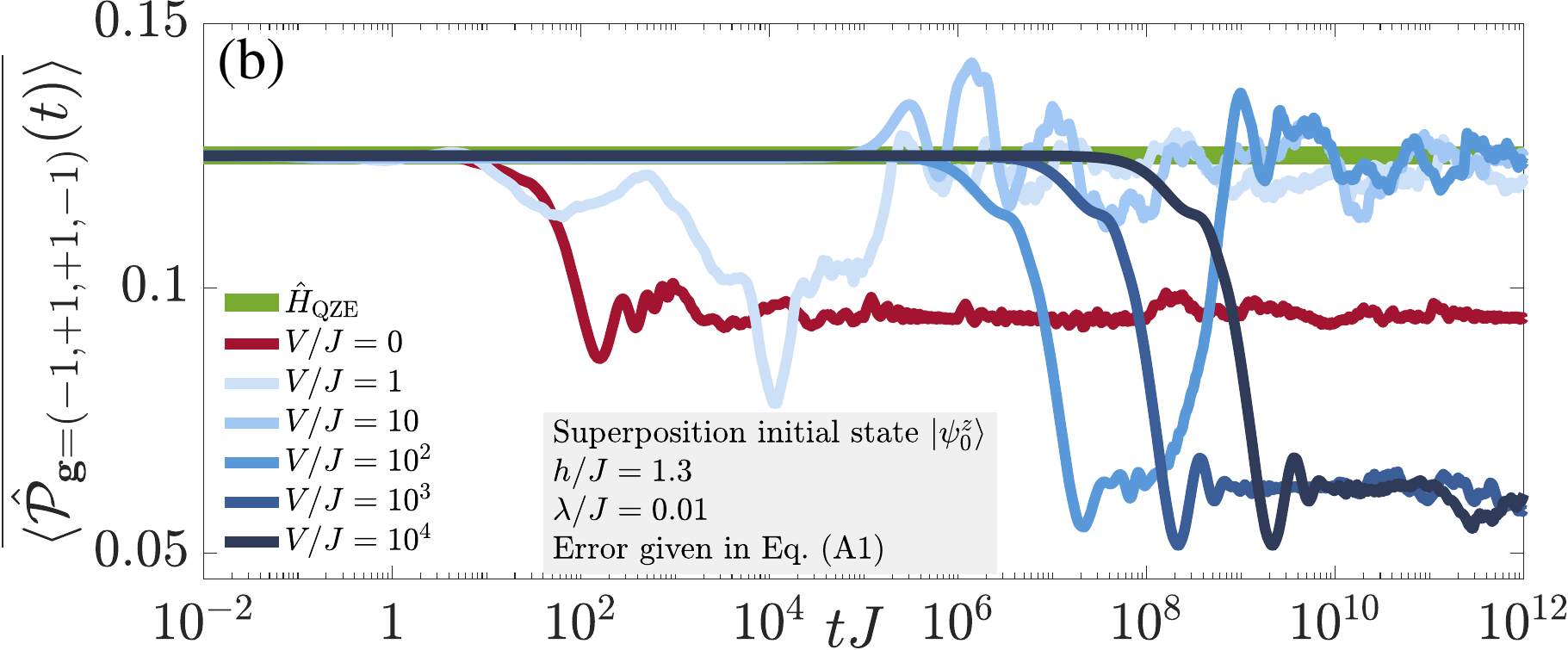}\\
	\vspace{0.1cm}
	\includegraphics[width=.48\textwidth]{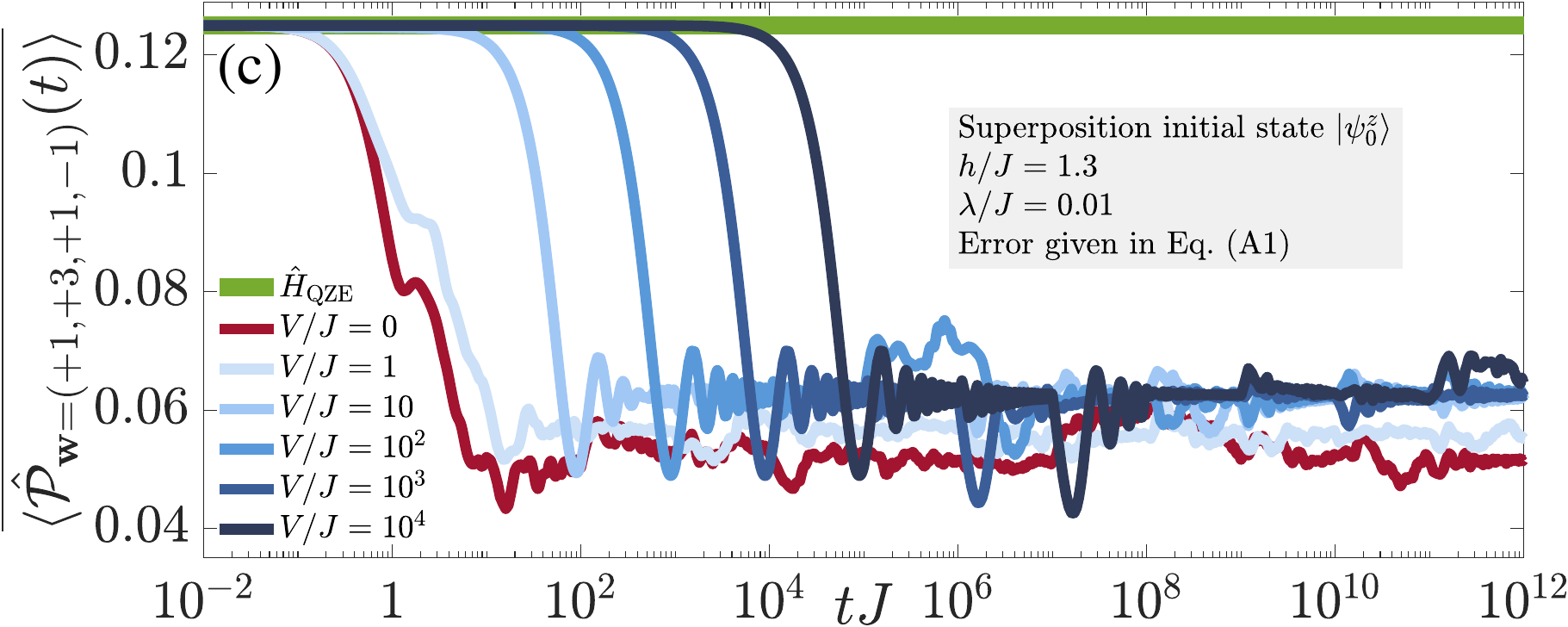}
	\caption{(Color online). Quench dynamics with the gauge-breaking term given in Eq.~\eqref{eq:H1simple}, which involves no gauge-invariant processes. (a) The time evolution of the imbalance is qualitatively identical to the case of the error~\eqref{eq:H1} of Fig.~\ref{fig:nonGI_imbalance}(b) except for the timescale of the second prethermal plateau, which ends at $t\propto V/\lambda^2$, in agreement with the analytic earliest-time prediction, rather than $t\propto V^2/(\lambda^2 J)$ as in the case of the error~\eqref{eq:H1}. (b) Same as Fig.~\ref{fig:projectors}(a) but for the error given in Eq.~\eqref{eq:H1simple}. The same qualitative difference occurs as in the second plateau of the imbalance, with $\langle\hat{\mathcal{P}}_\mathbf{g}\rangle$ deviating from its initial value at a timescale $\propto V/\lambda^2$ rather than $\propto V^2/(\lambda^2 J)$. This makes sense as the second prethermal plateau of the imbalance signifies an effective theory with only the $\mathbb{Z}_2$ gauge symmetry conserved. (c) Same as Fig.~\ref{fig:projectors}(c) but with the gauge-breaking term~\eqref{eq:H1simple}. The result is qualitatively unchanged since $\langle\hat{\mathcal{P}}_\mathbf{w}\rangle$ is only weakly dependent on $\lambda\hat{H}_1$, since $\hat{H}_0$ is the main term that breaks the enhanced local symmetry associated with $\hat{W}_j$.}
	\label{fig:SimpleError} 
\end{figure}

Let us simplify the error term in Eq.~\eqref{eq:H1} by removing from it the terms $\propto\eta_{1\ldots4}$, leaving us with the error term
\begin{align}\label{eq:H1simple}
	\lambda\hat{H}_1=\,\lambda\sum_{j=1}^{L}\big(\hat{a}_j^\dagger\hat{a}_{j+1}+\hat{a}_j\hat{a}_{j+1}^\dagger+\hat{\tau}^z_{j,j+1}\big),
\end{align}
In this case, the timescale at which $\langle\hat{\mathcal{P}}_\mathbf{g}\rangle$ deviates from its initial value is $\propto V/\lambda^2$, as shown in Fig.~\ref{fig:SimpleError} for the imbalance, and the projectors onto the superselection sectors $\mathbf{g}$ and $\mathbf{w}$. Once again, we see by comparing Fig.~\ref{fig:SimpleError}(a) and Fig.~\ref{fig:SimpleError}(b) that the second plateau of the imbalance begins to thermalize towards zero around the same time $\langle\hat{\mathcal{P}}_\mathbf{g}\rangle$ leaves its initial value, which signifies that the dynamics of this plateau is effectively under a renormalized theory with only the $\mathbb{Z}_2$ gauge symmetry preserved. Comparing Fig.~\ref{fig:SimpleError}(a) and Fig.~\ref{fig:SimpleError}(c), we find that the imbalance dynamics leaves its first plateau, which is well approximated by $\hat{H}_\mathrm{QZE}$ up to a timescale $V/J^2$, around the time $\langle\hat{\mathcal{P}}_\mathbf{w}\rangle$ deviates from its initial value. This is because $\hat{H}_\mathrm{QZE}$ hosts an enhanced local symmetry associated with $\hat{W}_j$.

It is worth noting about Fig.~\ref{fig:SimpleError} that the error involved, given in Eq.~\eqref{eq:H1simple}, involves no gauge-invariant processes, unlike the error of Eq.~\eqref{eq:H1}. This may explain why the second plateau, i.e., the one describing a renormalized $\mathbb{Z}_2$ gauge theory, is longer-lived in the case of error~\eqref{eq:H1} compared to that of error~\eqref{eq:H1simple}. Note that the timescale of the first plateau is largely independent of the nature of $\lambda\hat{H}_1$ (at least for small $\lambda$), and seems to be mostly dependent on $\hat{H}_0$ and the LPG protection strength.

\begin{figure*}[t!]
	\centering
	\includegraphics[width=.7\textwidth]{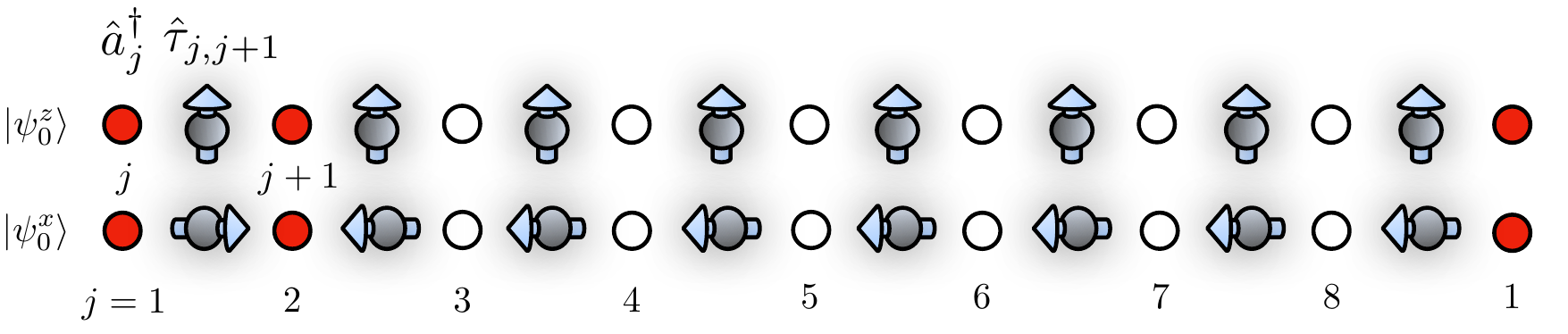}\\
	\vspace{0.1cm}
	\includegraphics[width=.48\textwidth]{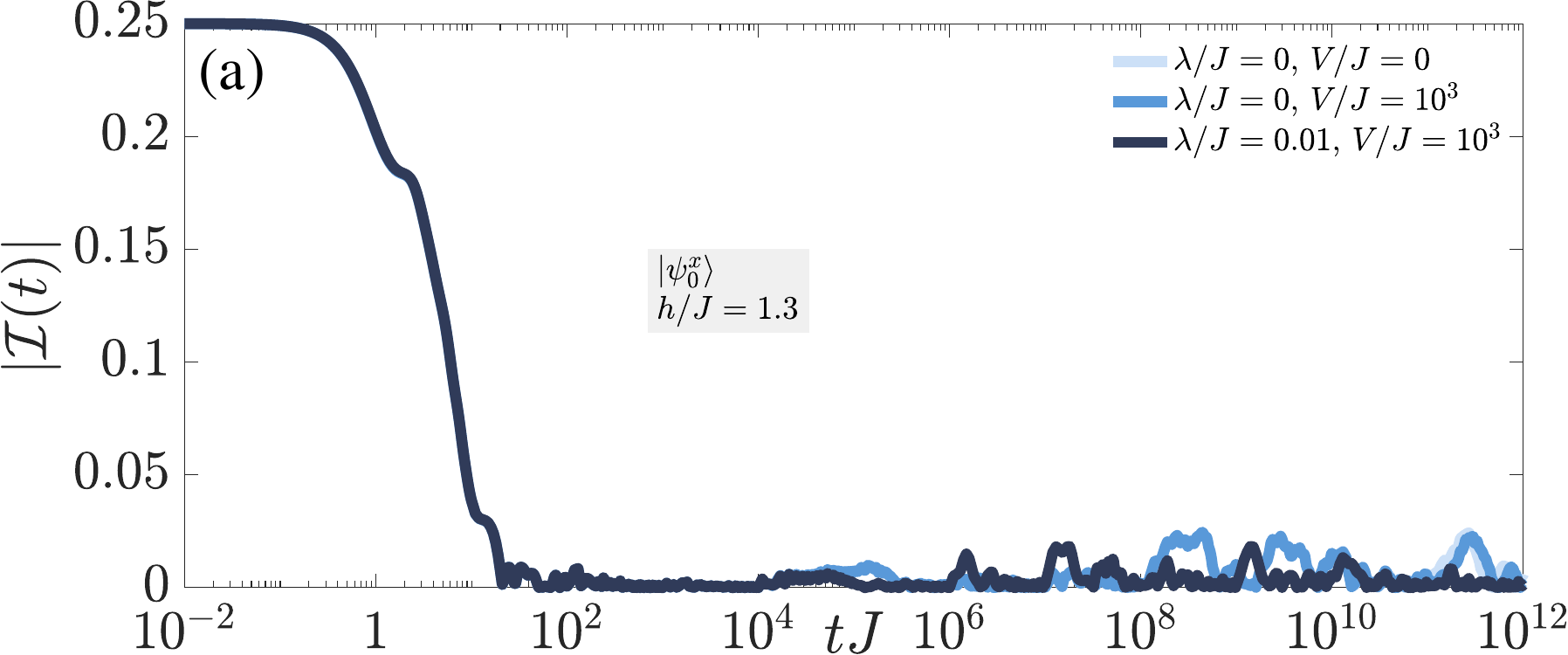}\quad
	\includegraphics[width=.48\textwidth]{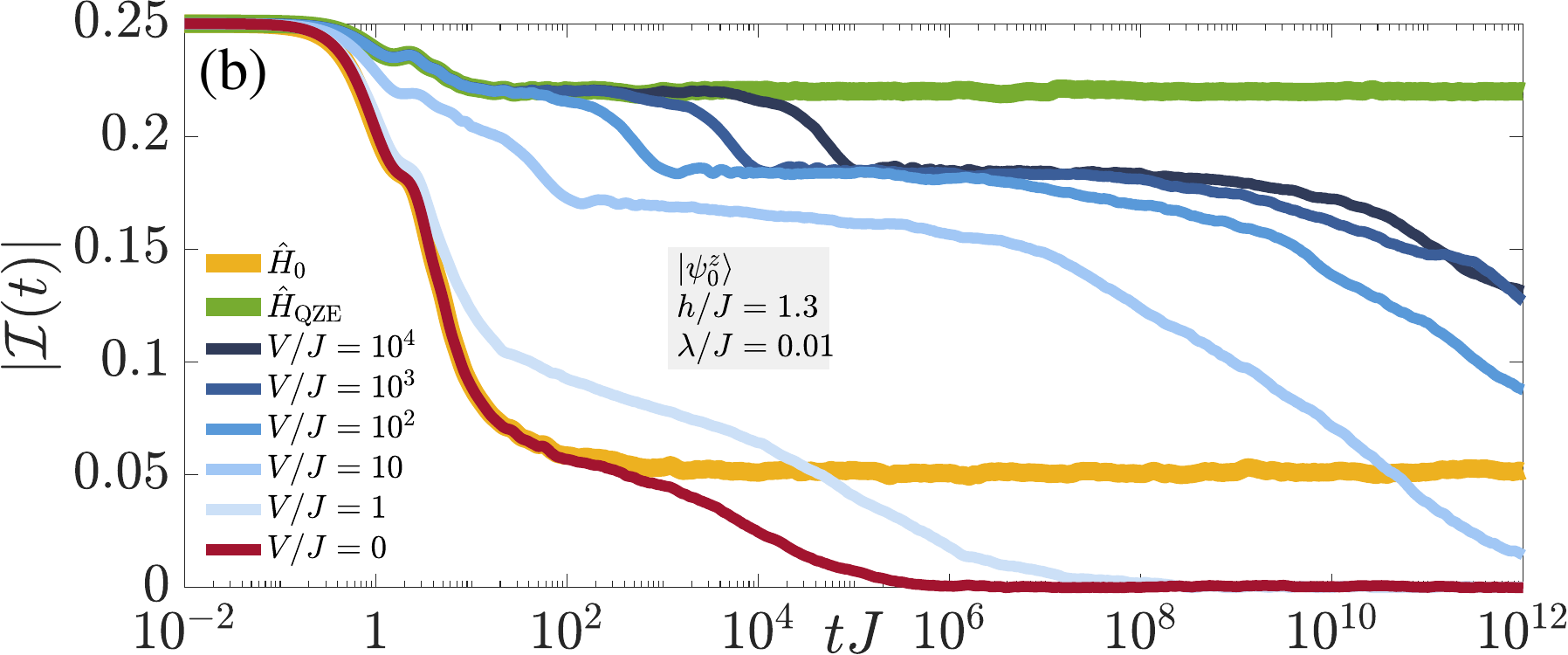}
	\caption{(Color online). (a) Same as Fig.~\ref{fig:GI_imbalance} but for a system with $L=8$ matter sites and $L=8$ gauge links at quarter filling in the bosons, with the gauge-invariant initial state $\ket{\psi^x_0}$ given on top. Disorder-free localization is absent and the system thermalizes with a vanishing imbalance regardless of the values of $\lambda$ and $V$. (b) Same as Fig.~\ref{fig:nonGI_imbalance}(b) but for a system with $L=8$ matter sites and $L=8$ gauge links at quarter filling in the bosons, with the gauge-noninvariant superposition initial state $\ket{\psi^z_0}$ given on top. LPG protection stabilizes disorder-free localization, where the dynamics at sufficiently large $V$ is faithfully reproduced by an emergent gauge theory $\hat{H}_\mathrm{QZE}=\sum_\mathbf{w}\hat{\mathcal{P}}_\mathbf{w}\big(\hat{H}_0+\lambda\hat{H}_1\big)\hat{\mathcal{P}}_\mathbf{w}$, which hosts an enhanced local symmetry associated with the LPG and that includes the original $\mathbb{Z}_2$ gauge symmetry, up to a timescale $\propto V/J^2$. After this timescale, a renormalized gauge theory with only the original $\mathbb{Z}_2$ gauge symmetry emerges up to timescales polynomial in $V$, after which thermalization occurs and the imbalance vanishes. }
	\label{fig:imbalance_L8} 
\end{figure*}

\subsection{Dependence on system size}
From the concept of quantum Zeno subspaces, we have been able to derive in Sec.~\ref{sec:QZE} an emergent gauge theory $\hat{H}_\mathrm{QZE}$ with an enhanced local symmetry associated with the pseudogenerator $\hat{W}_j$ that faithfully reproduces the dynamics of the faulty theory up to an earliest timescale $\propto V/(V_0 L)^2$. Our exact diagonalization results have confirmed this prediction and even exceeded it in certain cases for a fixed value of $L$. However, the question remains as to whether increasing the system size will quantitatively reduce this timescale at a given value of $V$ and $\lambda$.

To answer this question, let us double our system size, while keeping only two bosons on the lattice for reasons of numerical overhead. If we prepare our system in a gauge-invariant state $\ket{\psi^x_0}$, shown in Fig.~\ref{fig:imbalance_L8}, and quench with the faulty theory $\hat{H}=\hat{H}_0+\lambda\hat{H}_1+V\hat{H}_W$ with $\lambda\hat{H}_1$ given in Eq.~\eqref{eq:H1}, we observe no disorder-free localization, and the system thermalizes with a vanishing imbalance regardless of the values of $\lambda$ and $V$; see Fig.~\ref{fig:imbalance_L8}(a). On the other hand, if once again we prepare our system in a superposition of superselection sectors $\mathbf{g}$, then quenching the corresponding state $\ket{\psi^z_0}$ shown in Fig.~\ref{fig:imbalance_L8}(b) by $\hat{H}_0$ will lead to disorder-free localization that persists for all accessible times, and there is roughly the same ratio of memory retention as in the case of $L=4$ matter sites. Upon adding gauge-breaking errors, disorder-free localization is compromised and the system thermalizes with a monotonous decay to zero, as shown in Fig.~\ref{fig:imbalance_L8}(b). However, once LPG protection is turned on, disorder-free localization is restored, and at sufficiently large $V$ we find that the dynamics of the imbalance is faithfully reproduced by the emergent gauge theory $\hat{H}_\mathrm{QZE}$ up to a timescale $\propto V/J^2$. We see little dependence on system size, when comparing to Fig.~\ref{fig:nonGI_imbalance}(b), in the quantitative value of this timescale for a given value of $V$. This is encouraging for future large-scale experiments on disorder-free localization.

\section{Thermal ensembles}\label{app:thermal}
Due to the spatial homogeneity and translation-invariance of our system, it is intuitive to expect that the imbalance will thermalize to zero in case thermalization does take place. To check this, we look at the prediction due to both the microcanonical and canonical ensembles. 

The microcanonical ensemble $\hat{\rho}_\mathrm{ME}$ is constructed as follows. Let $\ket{E_n}$ be the eigenstates with eigenenergies $E_n$ of the quench Hamiltonian $\hat{H}$. For an initial state $\ket{\psi_0}$, the quench energy is $E_\mathrm{quench}=\bra{\psi_0}\hat{H}\ket{\psi_0}$. Then we can write
\begin{align}\label{eq:ME}
    \hat{\rho}_\mathrm{ME}=\frac{1}{\mathcal{N}_{\Delta E}}\sum_{ E_n\in\mathcal{W}}\ket{E_n}\bra{E_n},
\end{align}
where $\mathcal{N}_{\Delta E}$ is the number of eigenstates $\ket{E_n}$ in the energy window $\mathcal{W}=[E_\mathrm{quench}-\Delta E/2,E_\mathrm{quench}+\Delta E/2]$. In our code, we have set $\Delta E=0.1 J$, but we find that our results are not sensitive to its exact value.

The thermal canonical ensemble is given by
\begin{align}\label{eq:CE}
    \hat{\rho}_\mathrm{CE}=\frac{e^{-\beta\hat{H}}}{\mathcal{Z}},
\end{align}
where $\mathcal{Z}=\Tr\big\{e^{-\beta\hat{H}}\big\}$ is the partition function. The only unknown in Eq.~\eqref{eq:CE} is the inverse temperature $\beta$, which can be determined using, e.g., Newton's method to solve
\begin{align}
    \bra{\psi_0}\hat{H}\ket{\psi_0}=\Tr\big\{\hat{\rho}_\mathrm{CE}\hat{H}\big\},
\end{align}
which states that the initial quench energy is conserved in the unitary dynamics, and that if the system thermalizes, the canonical ensemble should correctly predict this energy.

We find that if our initial state is $\ket{\psi_0}=\ket{\psi^{x,z}_0}$ (see Fig.~\ref{fig:InitialStates}), then $\Tr\big\{\hat{\rho}_\mathrm{ME}\hat{\mathcal{I}}\big\}=\Tr\big\{\hat{\rho}_\mathrm{CE}\hat{\mathcal{I}}\big\}=0$, where $\hat{\mathcal{I}}=\sum_{j=1}^Lp_j\hat{n}_j/L$ for generic values of $\lambda$ and $V$. This agrees with the unitary dynamics we calculate in exact diagonalization for the case of the initial state $\ket{\psi^x_0}$, which thermalizes, but not $\ket{\psi^z_0}$, which leads to disorder-free localization.

\section{Linear protection in the local generator}\label{app:full}

\begin{figure}[t!]
	\centering
	\includegraphics[width=.48\textwidth]{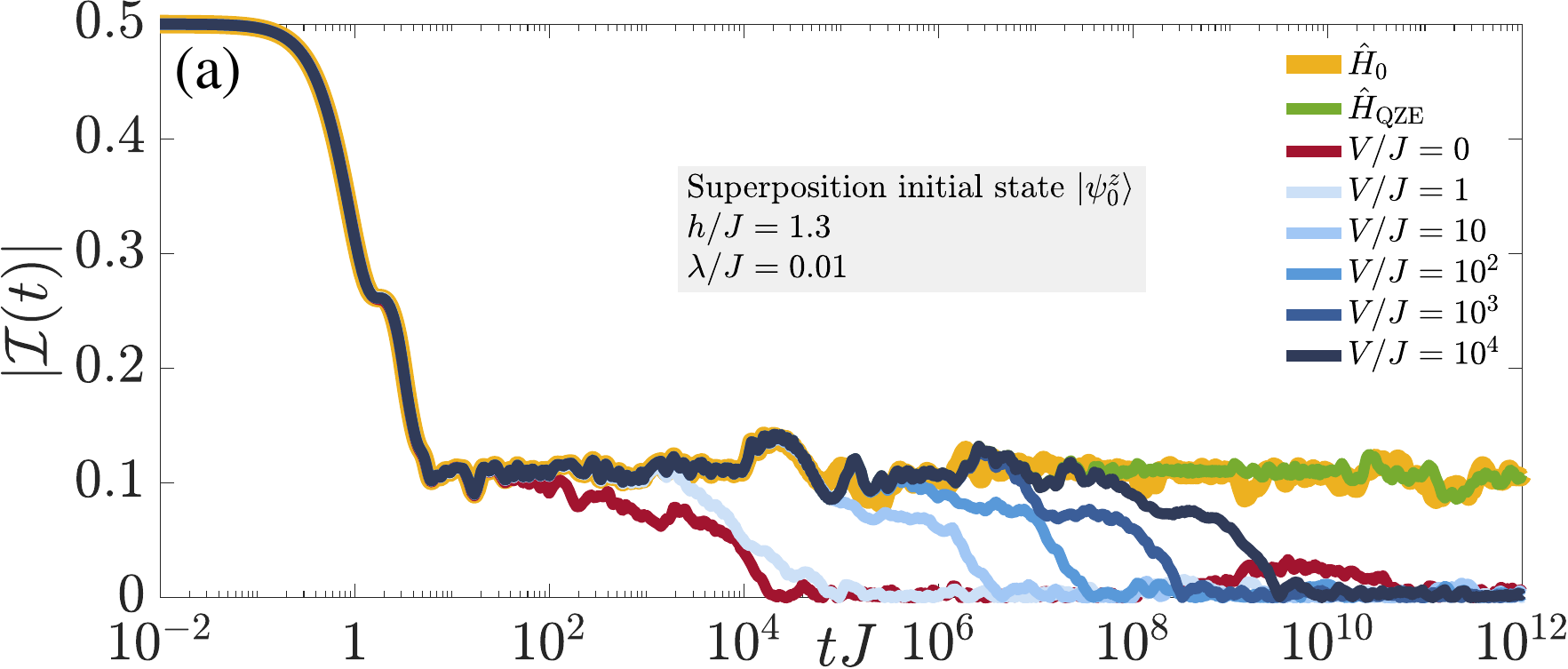}
	\caption{(Color online). Quench dynamics of the imbalance~\eqref{eq:imbalance} under $\hat{H}_0+\lambda\hat{H}_1+V\hat{H}_G$ with the error term~\eqref{eq:H1} and $c_j=[6(-1)^j+5]/11$. The protection term $V\hat{H}_G=V\sum_jc_j\hat{G}_j$ does not dynamically induce any new symmetries, but rather only protects the original $\mathbb{Z}_2$ symmetry of $\hat{H}_0$ through quantum Zeno dynamics.}
	\label{fig:HG} 
\end{figure}
The principle of quantum Zeno subspaces will also work when employing the actual local generator $\hat{G}_j$ in the protection term
\begin{align}
    V\hat{H}_G=V\sum_jc_jG_j,
\end{align}
where we will again use $c_j=[6(-1)^j+5]/11$. Let us now quench $\ket{\psi^z_0}$ with the faulty theory $\hat{H}=\hat{H}_0+\lambda\hat{H}_1+V\hat{H}_G$, with $\lambda \hat{H}_1$ given in Eq.~\eqref{eq:H1}, and calculate the ensuing dynamics of the imbalance, shown in Fig.~\ref{fig:HG}. The protection works remarkably well, except it is now fundamentally different from the LPG protection in that it cannot induce an enhanced local symmetry. The emergent gauge theory in this case is $\hat{H}_\mathrm{QZE}=\hat{H}_0+\lambda\sum_\mathbf{g}\hat{\mathcal{P}}_\mathbf{g}\hat{H}_1\hat{\mathcal{P}}_\mathbf{g}$ \cite{Halimeh2021stabilizingDFL}, and it reproduces the dynamics under $\hat{H}$ up to times polynomial in $V$, in accordance with the quantum Zeno effect. This effective gauge theory has only the original $\mathbb{Z}_2$ gauge symmetry due to $\hat{H}_0$. As such, $V\hat{H}_G$ only stabilizes disorder-free localization but does not enhance it (see Fig.~\ref{fig:HG}). Such a protection scheme has also been demonstrated to reliably stabilize disorder-free localization in spin-$S$ $\mathrm{U}(1)$ quantum link models \cite{Halimeh2021stabilizingDFL}.

Despite the theoretical efficacy of such a protection term, the implementation of the actual generators $\hat{G}_j$ in a quantum synthetic matter setup is quite challenging as they involve multi-species three-body terms, which are at least as difficult to implement as $\hat{H}_0$. The local pseudogenerator $\hat{W}_j$, on the other hand, comprises at most single-species two-body terms, which are easier to implement than the ideal gauge theory itself \cite{Halimeh2021stabilizing}.

\bibliography{DisFreeLoc_biblio}
\end{document}